\documentclass[sigconf]{acmart}
\AtBeginDocument{%
  }

\usepackage{algorithm}
\usepackage{algorithmic}

\usepackage{amsmath,amsfonts,bm}

\def\eqref#1{equation~\ref{#1}}

\def\1{\bm{1}}

\def\rve{{\mathbf{e}}}

\def\rvr{{\mathbf{r}}}

\def\rvt{{\mathbf{t}}}

\def\rvv{{\mathbf{v}}}

\def\rvx{{\mathbf{x}}}
\def\rvy{{\mathbf{y}}}
\def\rvz{{\mathbf{z}}}

\def\rmA{{\mathbf{A}}}

\def\rmC{{\mathbf{C}}}

\def\rmE{{\mathbf{E}}}

\def\rmI{{\mathbf{I}}}

\def\rmO{{\mathbf{O}}}

\def\rmR{{\mathbf{R}}}
\def\rmS{{\mathbf{S}}}
\def\rmT{{\mathbf{T}}}

\def\rmV{{\mathbf{V}}}

\DeclareMathAlphabet{\mathsfit}{\encodingdefault}{\sfdefault}{m}{sl}
\SetMathAlphabet{\mathsfit}{bold}{\encodingdefault}{\sfdefault}{bx}{n}

\def\gD{{\mathcal{D}}}
\def\gE{{\mathcal{E}}}

\def\gG{{\mathcal{G}}}

\def\gL{{\mathcal{L}}}

\def\gN{{\mathcal{N}}}

\def\gT{{\mathcal{T}}}

\def\gV{{\mathcal{V}}}

\def\sD{{\mathbb{D}}}

\def\sI{{\mathbb{I}}}

\def\sR{{\mathbb{R}}}

\def\sT{{\mathbb{T}}}
\def\sU{{\mathbb{U}}}
\def\sV{{\mathbb{V}}}

\newcommand{\E}{\mathbb{E}}

\newcommand{\KL}{D_{\mathrm{KL}}}

\usepackage{newfloat}
\usepackage{listings}
\usepackage{multirow}
\usepackage{graphics}
\usepackage{subcaption}
\usepackage{caption}
\usepackage{float}
\usepackage{bbding}
\usepackage{pifont}
\usepackage{amsmath}
\usepackage{amsthm}
\usepackage{amsfonts}

\usepackage{amssymb}
\usepackage{mathtools}
\usepackage{booktabs}
\usepackage[normalem]{ulem}
\useunder{\uline}{\ul}{}
\theoremstyle{definition}

\begin{document}

\title{User-Aware Conditional Generative Total Correlation Learning for Multi-Modal Recommendation}

\author{Jing Du}
\affiliation{
    \institution{The University of New South Wales}
    \city{Sydney}
    \country{Australia}
}
\email{jing.du2@unsw.edu.au}

\author{Zesheng Ye}
\affiliation{
    \institution{University of Melbourne}
    \city{Melbourne}
    \country{Australia}
}
\email{zesheng.ye@unimelb.edu.au}

\author{Congbo Ma}
\affiliation{
    \institution{New York University Abu Dhabi}
    \city{Abu Dhabi}
    \country{UAE}
}
\email{cm7196@nyu.edu}

\author{Feng Liu}
\affiliation{
    \institution{University of Melbourne}
    \city{Melbourne}
    \country{Australia}
    }
\email{fengliu.ml@gmail.com}

\author{Flora Salim}
\affiliation{
    \institution{The University of New South Wales}
    \city{Sydney}
    \country{Australia}
    }
\email{flora.salim@unsw.edu.au}

\renewcommand{\shortauthors}{Jing Du et al.}

\begin{abstract}
    {\em Multi-modal recommendation} (MMR) enriches item representations by introducing item content, e.g., visual and textual descriptions, to improve upon interaction-only recommenders.
    The success of MMR hinges on aligning these content modalities with user preferences derived from interaction data, yet dominant practices based on disentangling modality-invariant preference-driving signals from modality-specific preference-irrelevant noises are flawed.
    First, they assume a {\em one-size-fits-all} relevance of item content to user preferences for all users, which contradicts the user-conditional fact of preferences.
    Second, they optimize {\em pairwise} contrastive losses separately toward cross-modal alignment, systematically ignoring higher-order dependencies inherent when multiple content modalities jointly influence user choices.
    In this paper, we introduce GTC, a conditional \underline{\textit{G}}enerative \underline{\textit{T}}otal \underline{\textit{C}}orrelation learning framework.
    We employ an interaction-guided diffusion model to perform {\em user-aware} content feature filtering, preserving only personalized features relevant to each individual user.
    Furthermore, to capture complete cross-modal dependencies, we optimize a tractable lower bound of the total correlation of item representations across all modalities.
    Experiments on standard MMR benchmarks show GTC consistently outperforms state-of-the-art, with gains of up to 28.30\% in NDCG@5.
    Ablation studies validate both conditional preference-driven feature filtering and total correlation optimization, confirming the ability of GTC to model user-conditional relationships in MMR tasks.
    The code is available at: \url{https://github.com/jingdu-cs/GTC}.
\end{abstract}



\maketitle

\section{Introduction}
Recommender systems have evolved beyond collaborative filtering to embrace {\em multi-modal recommendation} (MMR), where diverse and rich content information, such as product images and textual descriptions, is used to enrich item representations by providing an explicit view of item characteristics that complements user preferences derived from sparse user-item interactions~\citep{liu2024multimodal}.
When these information-rich content modalities are {\em effectively aligned} with interaction data modality, MMR promises more comprehensive user preferences grounded in semantically-meaningful item attributes, thus improving recommendation performance~\citep{wei2019mmgcn, wang2021dualgnn}.

\begin{figure}
    \centering
    \begin{subfigure}[b]{.445\linewidth}
        \centering
        \includegraphics[width=\textwidth]{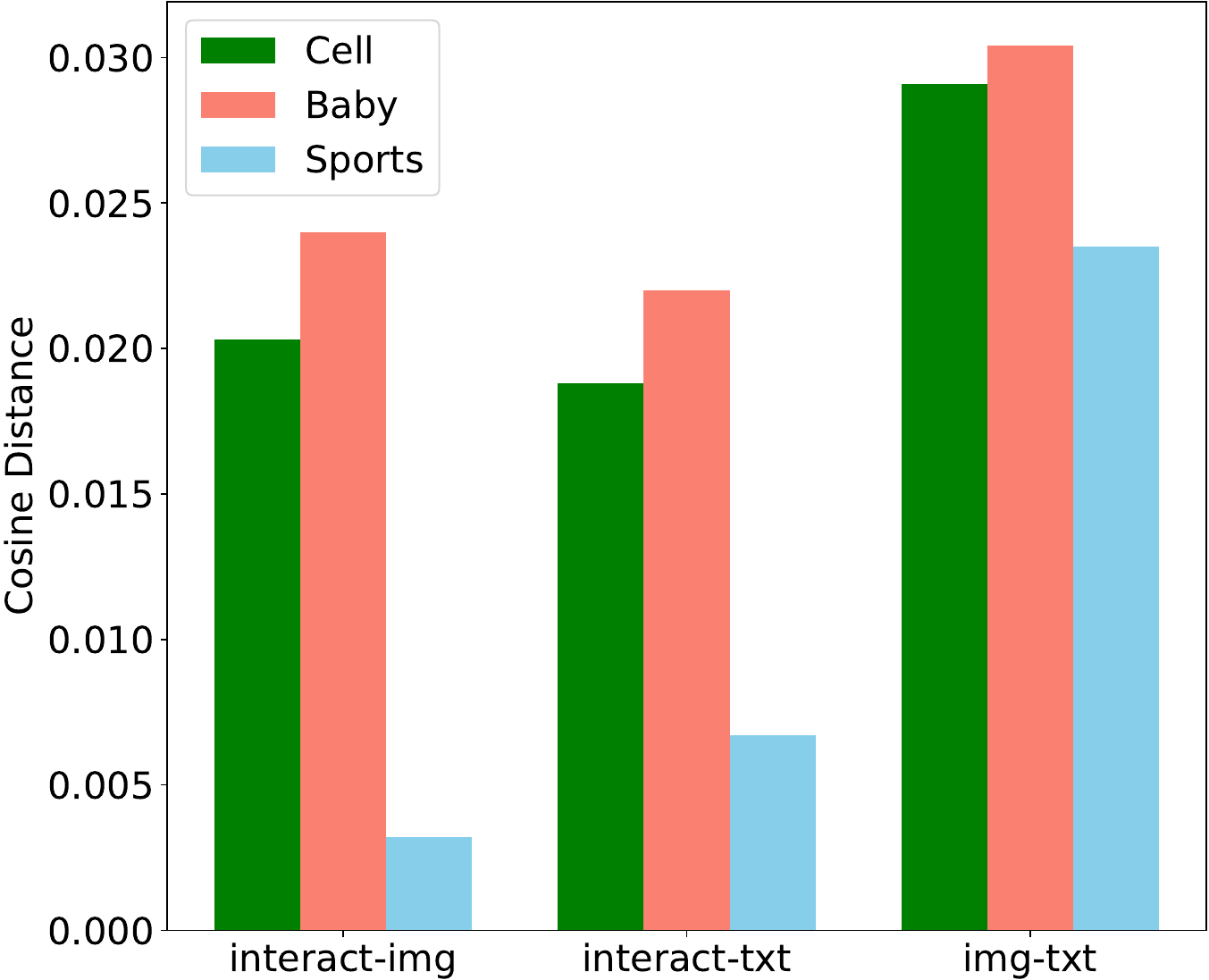}
        \label{fig:euclidean}
    \end{subfigure}
    \begin{subfigure}[b]{.44\linewidth}
        \centering
        \includegraphics[width=\textwidth]{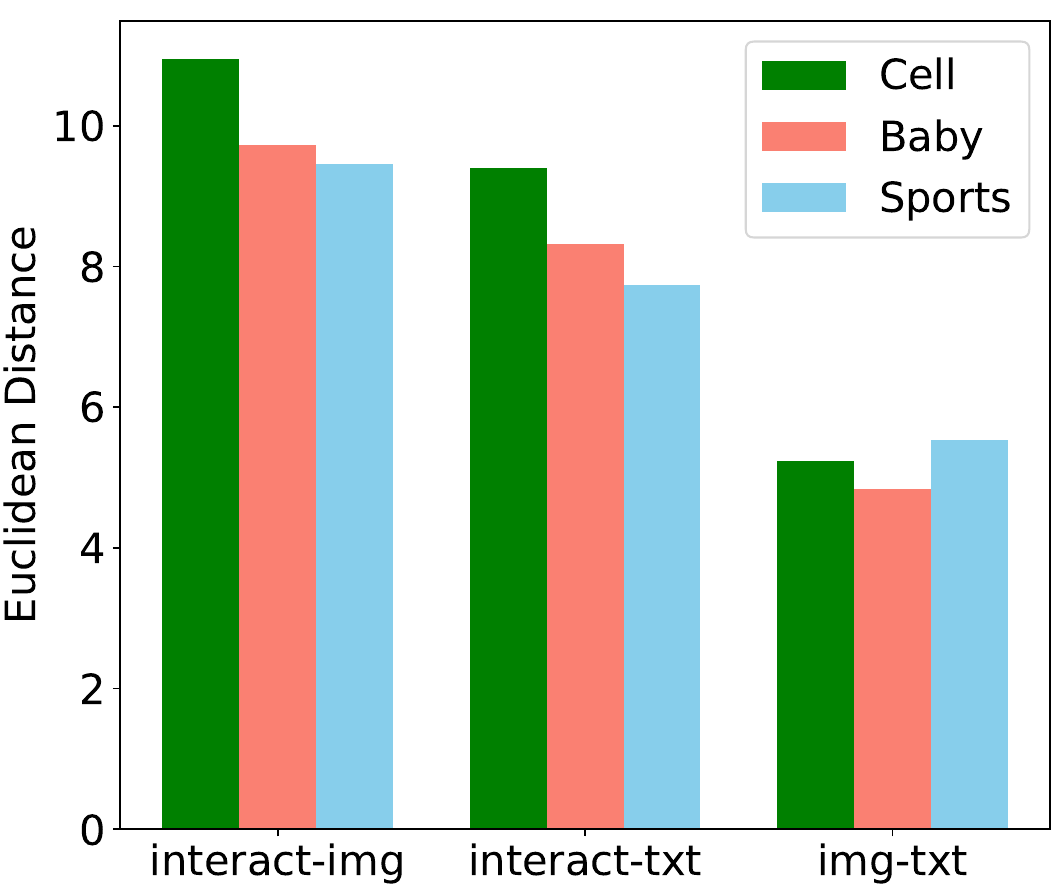}
        \label{fig:cosine}
    \end{subfigure}
        \vspace{-5mm}
    \caption{
        The representations from user-item interactions exhibit low cosine similarity (left) and high Euclidean distance (right) to representations from visual and textual content modalities, on Amazon Sports dataset. In contrast, visual and textual representations are more aligned, implying the gap between latent user preference and explicit item attributes.}
        \vspace{-5mm}
    \label{fig:toy_similarity}
\end{figure}

Yet, a tension exists inherently between these modalities.
Fig.~\ref{fig:toy_similarity} reveals a striking empirical observation: representations learned from user interactions diverge substantially from those extracted from visual and textual content, implying that the rich information in content modalities does {\em not} always align with the behavioral signals of user interactions holistically~\citep{kim2022mario}.
This divergence exposes a fundamental challenge in MMR.
Only partial features drive a user's decision to engaged with an item;
while the rest, from this user's perspective, are noise that do not help to reveal the true preference, reflected by user-item interactions.
As such, effective MMR models must be able to distinguish ``relevant'' preference-driving signals from ``irrelevant'' noises, filtering item features based on their relevance to user preference to avoid performance degradation~\citep{zhou2023enhancing}.

The predominant response is multi-modal {\em Disentangled Representation Learning}~(DRL), which attempts to isolate a ``modality-invariant'' preference-driving component of the item representation from ``modality-specific'' ones~\citep{liu2022multi, xu2024cmclrec}, and optimize for pairwise contrastive losses to align modalities thereafter, assuming that core preference, e.g., a T-shirt's appeals might be ``comfortable'', remain consistent across modalities and users; whilst noisy attributes, e.g., the use of words like ``must-have'', may vary instead~\citep{zhou2023bootstrap, cao2022cross}.

\begin{figure}
    \centering
    \includegraphics[width=.93\linewidth]{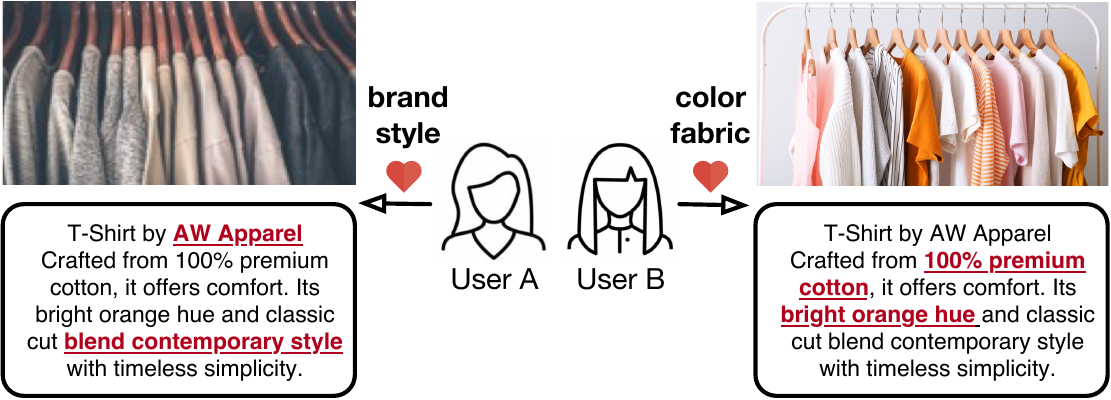}
    \caption{
        The user-conditional nature of ``appealing'' feature relevance.
        The item on the right appeals to a user who likes ``orange'' and ``cotton'', while a different user is more interested in the brand (``AW Apparel'') and style, for whom color and fabric are less important. Thus, what constitutes a preference-driving signal is not only determined by the item attributes themselves, but also by individual user preference.}
    \vspace{-5mm}
    \label{fig:example}
\end{figure}

However, the prevailing DRL practices are both {\bf impractical} and {\bf suboptimal} (detailed in Sec.~\ref{sec:prior_limitation}).
One issue is that they enforce a one-size-fits-all separation of item representations over all users.
This implicitly assumes the relevance of a feature to user preferences is {\em fixed} and {\em universal}, when in fact it is inherently {\em conditional on each user}.
A user may prioritize a T-shirt's ``orange'' color and ``cotton'' fabric, while another cares only about the brand (Fig. \ref{fig:example}).
By doing so, DRL methods may discard features relevant to some users but not others, failing to capture the entangled nature of item attributes and user-specific behaviors~\citep{cai2025learning}.

Second, existing MMR practices optimize pairwise contrastive losses independently---aligning interaction-textual, interaction-visual, and visual-textual modalities in parallel~\citep{zhou2023bootstrap, tao2022self}.
This is incomplete when {\em more than two modalities} are involved, as it cannot model interdependencies that emerge when modalities interact {\em collectively}.
Consider a user who might engage with a product only when both its visual appeal and textual description align with the interest, neither modality alone would drive the interaction.
Pairwise alignment treats visual and textual signals as independent factors, failing to model this joint dependency, resulting in suboptimal alignment.

Such limitations, i.e., the flawed assumption of universal content feature relevance to user preferences and the incomplete modeling of cross-modal dependencies, call for an alternative MMR paradigm.
In Sec.~\ref{sec:methodology}, we propose a new framework called GTC that learns to \underline{\textit{G}}enerate interaction-guided item representations for content modalities and maximize cross-modal \underline{\textit{T}}otal \underline{\textit{C}}orrelation.
GTC is built upon two principles:
(1) since user-item interactions directly reflects user preference towards items, it leverages a diffusion model, guided by individual user interaction histories, and performs {\em user-conditional} preference filtering to denoise visual and textual features and preserve only those relevant to that specific user interactions;
(2) it pursues {\em holistic cross-modal alignment} and maximizes the {\em total correlation}~\citep{watanabe1960information}, which captures the complete dependency structure, including higher-order interactions that pairwise methods miss.

Concretely, following standard practices~\citep{wei2019mmgcn, zhou2023enhancing}, GTC first encodes user embedding and {\em three} modality-specific item embeddings from interaction, textual and visual contents, based on user-item bipartite graphs.
It then {\em refines} the visual and textual representations using an interaction-guided diffusion model, leveraging its strength in representation learning~\citep{yang2023diffusion}.
Next, it {\em aligns} the interaction with two user-aware content representations by maximizing a tractable lower bound on their total correlation.
Lastly, it {\em integrates} the interaction item embedding with aligned content features to form the final representation for recommendation.
Our contributions are:

\begin{itemize}
    \item User-aware generative filtering that tailors content features to indivudal user preferences, eliminating impractical universal feature relevance assumptions (Sec.~\ref{sec:content_feature_generation}).
    \item Total correlation maximization, the first attempt to model higher-order cross-modal dependencies for MMR, mitigating issues of separate pairwise alignments (Sec.~\ref{sec:cross_modal_alignment_via_total_correlation}).
    \item A new framework that integrates these two principles to achieve a new state-of-the-art in MMR.
    \item Extensive validation across standard MMR benchmarks that confirm consistent effectiveness of GTC (Sec.~\ref{sec:experiments}).
\end{itemize}

\section{Background}
In this section, we first formulate the problem of MMR and contextualize previous work within this formulation.
We then pinpoint their limitations that motivate our approach.

\vspace{-2mm}
\subsection{Problem Formulation}
\paragraph{Recommendation.}
Let $\mathbb{U}=\{u_m\}_{m=0}^{|\mathbb{U}|}$ be the user set and $\mathbb{I}=\{i_n\}_{n=0}^{|\mathbb{I}|}$ be the item set.
Given an interaction matrix $\mathbf{R} \in \{0,1\}^{|\mathbb{U}| \times |\mathbb{I}|}$, where $R_{mn} = 1$ denotes an observed interaction exists between the $m$-th user and the $n$-th item.
These interactions are naturally represented as a user-item bipartite graph $\gG = \left< \sU, \sI, \gE \right>$, where an edge $(u_m, i_n) \in \gE$ exists if and only if $R_{mn} = 1$~\citep{cao2022cross,du2023distributional}.
The goal of recommendation is to learn a model parameterized by $\Theta$ that maps a user $u_m$ and an item $i_n$ to a shared latent representation space $\rve_m = h_u(\mathbf{R}; \Theta) \in \sR^d$ and $\rve_n = g_i(\mathbf{R}; \Theta) \in \sR^d$, by optimizing $\Theta$ to minimize a ranking-based loss function, e.g., Bayesian Personalized Ranking (BPR) loss~\citep{rendle2012bpr}, defined over a scoring function, typically the dot product $s(u_m, i_n) = \rve_m^\top \rve_n$.
Such that the scoring function $s(u_{m^{\prime}}, \cdot)$ can rank unseen items for each user $u_{m^{\prime}}$.

\vspace{-2mm}
\paragraph{Multi-modal Recommendation.}
MMR extends this setup by incorporating heterogeneous item content modalities.
We consider each $i_n$ has accompanying item image and textual description, pre-processed via pre-trained encoders, e.g., a convolutional neural network~\citep{he2016deep} for visual data and a Transformer~\cite{reimers2019sentence} for textual data, leading to visual features $\sV = \{ \rvv_i \in \sR^{d_v} \}_{i \in \sI}$ and textual features $\sT = \{ \rvt_i \in \sR^{d_t} \}_{i \in \sI}$ over all items.
Notice that the item representation map becomes $\rve_n = g_i(\rvv_i, \rvt_i, \mathbf{R}; \Theta)$ by integrating both visual and textual features, in addition to the interaction data.
The crucial design challenge in MMR, which has driven the evolution of the field, is how to design this function to effectively combine signals from the interaction modality with those from content modalities.
This paper categorizes them into two paradigms.

\vspace{-2mm}
\subsection{Previous Paradigms in MMR}
\paragraph{Direct Fusion.}
Early methods focused on learning representations for each modality and fusing them.
This began with direct concatenation~\citep{wei2019mmgcn} and evolved to using attention that assign weights to each modality, recognizing their unequal contributions to user preference~\citep{wang2021dualgnn, kim2022mario}.
A primary challenge with this paradigm is that content modalities are often noisy (i.e., features irrelevant to user preference) that can degrade performance when fused indiscriminately.
While later methods leveraged denoising steps like user-item interaction graph pruning~\citep{wei2020graph} and spectral filtering~\citep{zhou2023tale, SMORE}, they operate on a {\em user-agnostic} basis, either fusing all available information or filtering noise based on universal criteria, failing to model the fact that the feature relevance is often specific to individuals.

\paragraph{Disentanglement.}
The challenge of cleanly removing preference-irrelevant noise from content features spurred a paradigm shift towards {\em disentangled representation learning}~(DRL)~\citep{zhang2020content, ma2019learning, wang2022disentangled, an2025beyond}.
DRL methods aim to isolate a modality-invariant component (the assumed core preference-driving signal), from modality-specific components (the assumed noise)~\citep{du2023distributional, cao2022disencdr}.
This is often enforced via pairwise contrastive optimization, which aligns item representations from different modalities in a shared space~\citep{tao2022self, zhou2023bootstrap}.
Later studies focused on prompting statistical independence between signals and noise to facilitate disentanglement~\citep{guo2024lgmrec, lin2025contrastive}, sticking on fixed interaction-content relevance.

\subsection{Motivation of This Study}
\label{sec:prior_limitation}
Still, DRL inherits the user-agnostic flaw of direct fusion methods and even introduces new limitations.

\paragraph{Limitation of User-Agnostic Disentanglement.}
The core premise of DRL is that features can be {\em universally} categorized as ``relevant'' or ``irrelevant'' to user preference \cite{cao2022disencdr}.
However, this is misaligned with how preferences are formed.
Feature relevance is inherently user-conditional, as two users can have distinct preference patterns: a fashion-conscious person for whom {\em visual} aesthetics drive purchasing decisions, and a functionality-focused user who prioritizes {\em textual} technical specifications.
The enforcement of universal decomposition inevitably discards certain features that might be vital to specific users, creating a bottleneck affecting users whose preferences deviate from population-level patterns.
This highlights the need for user-aware preference filtering that can adaptively determine feature relevance of content, based on individual interaction patterns.
Such methods should preserve the benefits of noise reduction while maintaining sensitivity to user-specific preference.

\paragraph{Limitation of Pairwise Alignment.}
Also, the alignment strategy used in previous DRL methods is theoretically incomplete.
A holistic alignment requires capturing the complete statistical dependence among all modalities, a quantity measured by {\em total correlation}~\citep{watanabe1960information}.
Let $\rmS, \rmV, \rmT$ be the random variables for the interaction, visual, and textual representations of an item.
Following~\citep{saporta2025contrasting}, total correlation ${\rm TC}(\rmS, \rmV, \rmT)$ is defined as the KL-divergence between their joint distribution and the product of their marginals, decomposed as:
\begin{equation}\label{eq:tc_def}
    \begin{aligned}
        3 \; \cdot \; & {\rm TC} (\rmS, \rmV, \rmT)                                                                                                                                                                \\
                      & = 3 \cdot \KL(p(\rmS, \rmV, \rmT) \parallel p(\rmS) p(\rmV) p(\rmT))                                                                                                                       \\
                      & =  2 \cdot \underbrace{\left[ I(\rmS; \mathbf{V}) + I(\rmS; \mathbf{T}) + I(\mathbf{V}; \mathbf{T}) \right]}_{\text{pairwise alignment, captured by prior studies}}                        \\
                      & \quad + \underbrace{ I(\rmS; \mathbf{V} | \mathbf{T}) + I(\rmS; \mathbf{T} | \mathbf{V}) + I(\mathbf{V}; \mathbf{T} | \rmS) }_{\text{higher-order dependencies, missed by prior studies}},
    \end{aligned}
\end{equation}
where $I(\cdot; \cdot)$ is the mutual information between two random variables.
Unequivocally, pairwise contrastive alignment that maximizes a sum of pairwise mutual information between every two modalities, implemented in existing studies, is an incomplete proxy for total correlation.
They neglect higher-order dependencies, such as the information shared between interaction and text once the visual context is known, leading to a {\em suboptimal} alignment that fails to model the whole picture.

To overcome the limitations, we propose the principled GTC that departs from universal disentanglement assumptions to perform user-conditional feature filtering, and replaces incomplete pairwise alignment with an optimization towards total correlation\footnote{The primary obstacle to this is the intractability of optimizing total correlation directly, we will address this in Sec.~\ref{sec:cross_modal_alignment_via_total_correlation}.}.

\section{Methodology}\label{sec:methodology}
\subsection{Overview}

\begin{figure*}[ht]
    \centering
    \includegraphics[width=0.9\linewidth]{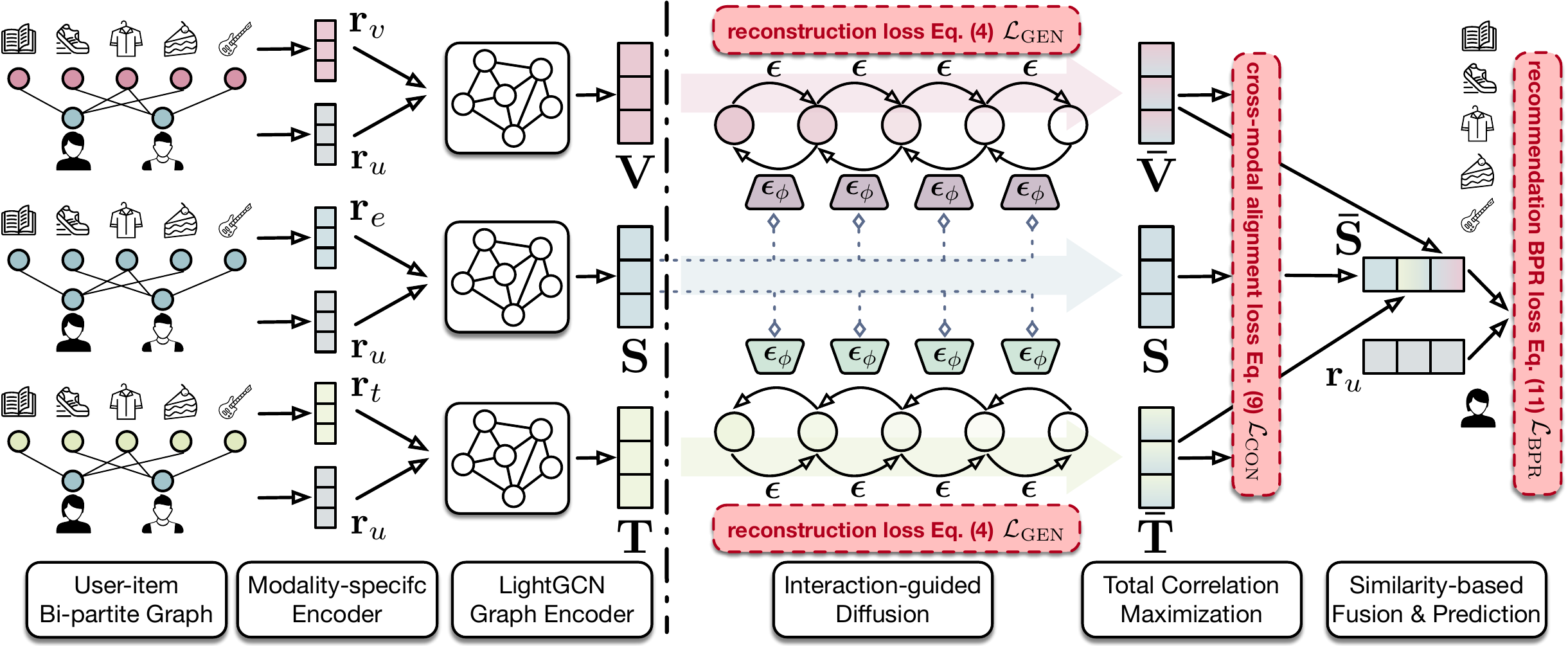}
    \caption{Overall illustration of the proposed GTC framework.
    }
    \label{fig:overall}
    \vspace{-5mm}
\end{figure*}

We now introduce the proposed framework for MMR, which learns to \underline{{\em G}}enerate user-aware item representations and maximizes cross-modal \underline{{\em T}}otal \underline{{\em C}}orrelation (GTC) with two design choices: (1) generating user-aware content features via user-item interaction guided diffusion (Sec.~\ref{sec:content_feature_generation}); and (2) maximizing (a tractable lower bound of) the total correlation for holistic cross-modal alignment (Sec.~\ref{sec:cross_modal_alignment_via_total_correlation}).
These aligned representations are then fused to perform the recommendation task (Sec.~\ref{sec:fused_representations_for_recommendation}).
Fig.~\ref{fig:overall} overviews the overall framework.


\vspace{-3mm}
\subsection{User-Aware Content Feature Generation}\label{sec:content_feature_generation}
\paragraph{Initial Embeddings.}
We begin by obtaining initial embeddings for all nodes in the bipartite graph $\gG$.
For the interaction modality, we randomly initialize a node embedding for each user $\rvr_u$ and item $\rvr_e$, capturing a latent collaborative space.
For content modalities, item features are captured by pre-trained encoders, denoted as $\rvr_v$ and $\rvr_t$ for visual and textual node features.
Since users do not have contents in our setup, their representations in the content modalities are given by the same initialized embeddings $\rvr_u$.
We then obtain modality-specific embeddings by propagating corresponding features through three parallel LightGCNs~\citep{he2020lightgcn} over the shared graph structure.
Let $\rmR_m$ be the input feature matrix for modality $m$ and $\rmE_m$ be the output embedding matrix, we have
\begin{equation}
    \rmE_m =  \text{LightGCN}(\bar{\rmA}, \rmR_m), \; \text{for } m \in \{ \gE, \gV, \gT \},
\end{equation}
where $\gE, \gV, \gT$ are the interaction, visual and textual spaces.
$\bar{\rmA}$ is the normalized adjacency matrix of the whole bipartite graph $\gG$.
From these outputs, we define three modality-specific embedding matrices: user and item $\rmS \coloneqq \rmE_\gE$ from interaction signals $\gE$, visual and textual $\rmV \coloneqq \rmE_\gV, \rmT \coloneqq \rmE_\gT$ from respective modalities.

\vspace{-2mm}
\paragraph{Interaction-Guided Denoising.}
The first core step of GTC is to refine the generic content embeddings $(\rmV, \rmT)$ to reflect individual user preferences.
To achieve this, we regard \textit{the interaction embeddings $\rmS$ as a reflection of user preference} and treat it as a user-specific guidance to denoise the content features $\rmV$ and $\rmT$.
Concretely, we learn an interaction-guided diffusion model over the content embedding spaces~\citep{yang2023diffusion}, which we apply independently to each initial content embedding matrix $\rmC_0 \in \{ \rmV, \rmT \}$.

The forward diffusion process defines a Markov chain that gradually injects Gaussian noise into the content embedding $\rmC_0$ over $T$ timesteps.
It ends up with $\rmC_{T}$ being a pure Gaussian, from which the state $\rmC_t$ at any timestep $t$ can be sampled in a closed form:
\begin{equation}\label{eq:forward}
\vspace{-1mm}
    \rmC_t = \sqrt{\bar{\alpha}_t}\rmC_0 + \sqrt{1-\bar{\alpha}_t}\boldsymbol{\epsilon},
\end{equation}
where $\boldsymbol{\epsilon} \sim \mathcal{N}(0, \rmI)$, and $\bar{\alpha}_t = \prod_{s=1}^t (1-\beta_s)$ is determined by a predefined linear variance schedule $\{\beta_t\}_{t=1}^T$.

The reverse process aims to invert Eq.~(\ref{eq:forward}) by modeling the conditional distribution $p_{\theta}(\rmC_{t-1}|\rmC_{t}, \rmS)$ parameterized by a neural network $\boldsymbol{\epsilon}_{\phi}$ that predicts the noise $\boldsymbol{\epsilon}$ added at timestep $t$ during the forward corruption process.
This denoising network is conditioned on the interaction embeddings $\rmS$, which encode user-specific interaction patterns.
Intuitively, conditioning on $\rmS$ encourages $\boldsymbol{\epsilon}_{\phi}$ to leverage user preference signals to determine which parts of the intermediate content signal are noise and should therefore be removed.
We instantiate $\boldsymbol{\epsilon}_{\phi}$ with a U-Net architecture~\citep{ronneberger2015u}.
To provide temporal context, we map timestep $t$ to a high-dimensional embedding ${\rm PE}(t)$ using sinusoidal positional encoding~\citep{vaswani2017attention}, which is injected into the network.
Following the simplified training objective \cite{ho2020denoising}, the network $\boldsymbol{\epsilon}_{\phi}$ is trained by minimizing the noise-prediction loss over timesteps $t = 1, \ldots, T$,
\begin{equation}\label{eq:gen_loss}
\gL_{\rm GEN}
= \E_{t, \rmC_0, \boldsymbol{\epsilon}} \left[
\left\lVert \boldsymbol{\epsilon} - \boldsymbol{\epsilon}_\phi\bigl(\rmC_t, t, \rmS\bigr) \right\rVert^2
\right].
\end{equation}
Note that Eq.~(\ref{eq:gen_loss}) applies to both $\rmV$ and $\rmT$ simultaneously.

\paragraph{Content Features Generation.}
After training, we generate the user-conditional content representations $\widetilde{\rmC}_0 \in \{\widetilde{\rmV}_0, \widetilde{\rmT}_0\}$ from the learned diffusion model.
We first sample the terminal state from a standard Gaussian $\rmC_T \sim \gN(\mathbf{0}, \rmI)$, and then iteratively denoise $\rmC_T$ back to $\rmC_0$ using the trained denoising network $\boldsymbol{\epsilon}_{\phi}$.
Concretely, for timesteps $t = T, \ldots, 1$, we update
\begin{equation}
\widetilde{\rmC}_{t-1}
= \frac{1}{\sqrt{\alpha_t}}
\left(
\widetilde{\rmC}_t
- \frac{1-\alpha_t}{\sqrt{1-\bar{\alpha}_t}},
\boldsymbol{\epsilon}_{\phi}(\widetilde{\rmC}_t, t, \rmS)
\right)
+ \sigma_t \rvz,
\end{equation}
where $\rvz \sim \mathcal{N}(\mathbf{0}, \rmI)$ is sampled from standard Gaussian, and $\sigma_t$ controls the stochasticity.
$\boldsymbol{\epsilon}_\phi$ is instantiated as a U‑Net architecture following~\citep{ronneberger2015u}.
We omit further architectural details for brevity.

\subsection{Modal Alignment via Total Correlation}\label{sec:cross_modal_alignment_via_total_correlation}
Having generated user-aware content representations, we align $\rmS$ with $\widetilde{\rmV}_0$ and $\widetilde{\rmT}_0$ by maximizing their total correlation ${\rm TC}(\rmS, \widetilde{\rmV}_0, \widetilde{\rmT}_0)$, which quantifies how strongly the triplet $(\rmS, \widetilde{\rmV}_0, \widetilde{\rmT}_0)$ are dependent.
As Eq.~(\ref{eq:tc_def}) establishes, maximizing ${\rm TC}(\rmS, \widetilde{\rmV}_0, \widetilde{\rmT}_0)$ is sufficient to simultaneously optimize both the pairwise alignment (e.g., $I(\rmS; \widetilde{\rmV}_0)$) and the higher-order dependencies (e.g., $I(\rmS; \widetilde{\rmV}_0 | \widetilde{\rmT}_0)$) that simpler pairwise objectives neglect~\citep{zhou2023bootstrap, tao2022self}.

For notational convenience, we hereinafter denote $\bar{\rmV}$ and $\bar{\rmT}$ the generated content representations $\widetilde{\rmV}_0$ and $\widetilde{\rmT}_0$.
Defined as the KL divergence between the joint distribution and the product of marginals~\cite{watanabe1960information}, 
$\rm{TC}(\rmS, \bar{\rmV}, \bar{\rmT})$ can be expressed as:
\begin{equation}\label{eq:tc_s_vbar_tbar}
    \begin{aligned}
        {\rm TC}(\rmS, \bar{\rmV}, \bar{\rmT}) & = \E_{\rmS, \bar{\rmV}, \bar{\rmT}} \left[ \log \frac{p(\rmS, \bar{\rmV}, \bar{\rmT})}{p(\rmS)p(\bar{\rmV})p(\bar{\rmT})} \right].
    \end{aligned}
\end{equation}
Since direct maximization of Eq.~(\ref{eq:tc_s_vbar_tbar}) is intractable, we follow~\cite{saporta2025contrasting} and optimize a tractable variational lower bound on Eq.~(\ref{eq:tc_s_vbar_tbar}) via the InfoNCE contrastive loss~\cite{oord2018representation}, aligning the modalities by distinguishing samples from the joint distribution $p(\rmS, \bar{\rmV}, \bar{\rmT})$ against those sampled from the product of marginals.


\paragraph{Multi-modal Contrastive Formulation.}
We thus formulate the alignment objective as a contrastive learning task over $M$ mini-batches of multi-modal samples from the underlying data distribution.

\noindent
\textbf{Sampling Strategy.}
For a batch of $N$ samples, we treat the matched triplet $(\rmS_i, \bar{\rmV}_i, \bar{\rmT}_i)$ as the {\em positive} anchor for $i = 1, \dots, N$ drawn from the joint distribution $p(\rmS, \bar{\rmV}, \bar{\rmT})$.
To construct {\em negative} samples that approximate the product of marginals, we {\em independently shuffle} the indices of the visual and textual representations within the batch, denoted by $\pi_v$ and $\pi_t$.
For each anchor $i$, we thus construct $N-1$ {\em negative} samples against the matched {\em positive} triplet $(\rmS_i, \bar{\rmV}_i, \bar{\rmT}_i)$.

\noindent
\textbf{Objective.}
We define the contrastive log-likelihood for anchor $\rmS$ as:
\begin{equation}
    \gL_{\rm CON}^{\rmS \to \bar{\rmV}, \bar{\rmT}} = \log \sum_{i=1}^{N} \frac{\exp h(\rmS_i, \bar{\rmV}_i, \bar{\rmT}_i)}{\sum_{j \neq i} \exp h(\rmS_i, \bar{\rmV}\pi_{v(j)}, \bar{\rmT}\pi_{v(j)})},
\end{equation}
where $h(\rvx, \rvy, \rvz) \triangleq \sum_{d=1}^D \rvx^{(d)} \rvy^{(d)} \rvz^{(d)}$ is a multilinear inner product.
Maximizing this objective corresponds to maximizing a lower bound on ${\rm TC}(\rmS, \bar{\rmV}, \bar{\rmT})$~\cite{watanabe1960information, saporta2025contrasting}:
\begin{equation}\label{eq:tc_bound}
    {\rm TC}(\rmS, \bar{\rmV}, \bar{\rmT}) \ge \log N + \E_{\gD} [\gL_{\rm CON}^{\rmS \to \bar{\rmV}, \bar{\rmT}}].
\end{equation}
Here, the expectation is taken over $M$ mini-batches of samples.
According to~\cite{oord2018representation}, this bound holds because the optimal $h^*$ implicitly estimates the log-density ratio defined within Eq.~(\ref{eq:tc_s_vbar_tbar}).

\noindent
\textbf{Symmetrization.}
As ${\rm TC}(\rmS, \bar{\rmV}, \bar{\rmT})$ is a symmetric measure over three modalities, while the single-anchor loss $\gL_{\rm CON}^{\rmS \to \bar{\rmV}, \bar{\rmT}}$ is not, we symmetrize the objective to cover all modal perspectives, such that
\begin{equation}\label{eq:contrastive_loss}
\gL_{\rm CON}
= \gL_{\rm CON}^{\rmS \to \bar{\rmV}, \bar{\rmT}}
+ \gL_{\rm CON}^{\bar{\rmV} \to \rmS, \bar{\rmT}}
+ \gL_{\rm CON}^{\bar{\rmT} \to \rmS, \bar{\rmV}}.
\end{equation}
By minimizing Eq.~(\ref{eq:contrastive_loss}), we encourage the model to capture higher-order dependencies among $\rmS$, $\bar{\rmV}$, and $\bar{\rmT}$ that are inaccessible to purely pairwise contrastive objectives, leading to a more comprehensive multi-modal alignment.

\subsection{Fused Representation for Recommendation}\label{sec:fused_representations_for_recommendation}
\paragraph{Similarity-based Fusion.}
With user-aware and aligned representations $\rmS, \bar{\rmV}, \bar{\rmT}$ available, we now integrate them into a coherent representation.
For each item (or user–item entity) with embeddings $\rmS_i$, $\bar{\rmV}_i$, and $\bar{\rmT}_i \in \mathbb{R}^D$, we first construct a fused content representation via element-wise interaction $\bar{\rmC}_f = \bar{\rmV} \odot \bar{\rmT}$.
We then use this representation to refine the original interaction embedding $\rmS$ using a similarity-based gating mechanism that dynamically controls the content contributions based on their relevance to the interaction patterns.
This yields a content-aware update to $\rmS$ as
\begin{equation}
\vspace{-2mm}
    \rmS_f = \alpha \cdot \bar{\rmC}_f = {\rm softmax} \left( \frac{ \rmS \cdot \bar{\rmC}_f / \tau }{ || \rmS || \cdot || \bar{\rmC}_f || } \right) \cdot \bar{\rmC}_f,
\end{equation}
where $\tau$ is a learnable temperature parameter.
To preserve the primary collaborative signal in $\rmS$ while allowing the content signal to enhance it in a similarity-adaptive manner, we obtain the final fused representation via a residual connection $\bar{\rmS} = \rmS + \rmS_f$.

\vspace{-2mm}
\paragraph{Recommendation.}
The final user and item representations are used to compute the ranking score $s(u, i)$ and predict the user's next interaction.
We optimize the interacted items to be ranked higher than un-interacted ones by minimizing the BPR loss~\citep{rendle2012bpr}:
\begin{equation}\label{eq:recom_loss}
    \gL_{\rm BPR} = \sum_{(u, i^{\rm po}, i^{\rm ne}) \in \sD} - \log \sigma (s(u, i^{\rm po}) - s(u, i^{\rm ne})),
\end{equation}
where $\sD$ is the set of training triplets, and $i^{\rm po}$ and $i^{\rm ne}$ are interacted and un-interacted items for user $u$. $\sigma(\cdot)$ is the sigmoid function.

\subsection{Training Objective}
The complete GTC framework is trained end-to-end by minimizing a composite objective function that combines recommendation loss (Eq.~(\ref{eq:recom_loss})), generation loss (Eq.~(\ref{eq:gen_loss})), and total correlation lower bound (Eq.~(\ref{eq:contrastive_loss})), with $\ell_2$ regularization applied to parameters $\Theta$, such that
\begin{equation}
\vspace{-1mm}
    \gL_{\rm GTC} = \gL_{\rm BPR} + \omega_1 \cdot \gL_{\rm GEN} + \omega_2 \cdot \gL_{\rm CON} + || \Theta ||^2,
\end{equation}
where $\omega_1$ and $\omega_2$ are hyperparameters that control the relative importance of the three components.

\begin{table*}
\vspace{-2mm}
    \caption{Overall Performance in Sports (up), Baby (middle), and Cell (down) dataset.
        For each comparison, the best-performing method is {\bf bolded}.
        \textit{Improved} indicates the percentage improvement of the proposed GTC over the \underline{runner-up model}.
        * denote statistically significant improvements, validated by a paired t-test at a significance level of $p < 0.05$ against the \underline{runner-up model}.}
    \label{tab: sportsresult}
    \centering
    \vspace{-2mm}
    \resizebox{0.85\linewidth}{!}{
    \begin{tabular}{cl|cccccc|cccc|cc}
    \toprule
    \multicolumn{2}{c|}{Models}                        & FREEDOM & SMORE  & GRCN   & LATTICE & BM3    & PGL    & SLMRec & LGMRec & MGCN   & DRAGON       & GTC             & Improved(\%) \\ \hline
    \multicolumn{1}{c|}{\multirow{4}{*}{NDCG}}   & @5  & 0.0196  & 0.0181 & 0.0245 & 0.0262  & 0.0267 & 0.0294 & 0.0291 & 0.0291 & 0.0314 & {\ul 0.0749} & \textbf{0.0961} & 28.30\%    \\
    \multicolumn{1}{c|}{}                        & @10 & 0.0257  & 0.0226 & 0.0316 & 0.0335  & 0.0343 & 0.0376 & 0.0368 & 0.0375 & 0.0399 & {\ul 0.0880} & \textbf{0.1106} & 25.68\%    \\
    \multicolumn{1}{c|}{}                        & @20 & 0.0327  & 0.0274 & 0.0394 & 0.0420  & 0.0430 & 0.0472 & 0.0451 & 0.0468 & 0.0496 & {\ul 0.1005} & \textbf{0.1242} & 23.58\%    \\
    \multicolumn{1}{c|}{}                        & @50 & 0.0437  & 0.0336 & 0.0510 & 0.0544  & 0.0552 & 0.0611 & 0.0564 & 0.0601 & 0.0636 & {\ul 0.1175} & \textbf{0.1411} & 20.09\%    \\ \hline
    \multicolumn{1}{c|}{\multirow{4}{*}{Recall}} & @5  & 0.0290  & 0.0268 & 0.0373 & 0.0392  & 0.0397 & 0.0441 & 0.0434 & 0.0442 & 0.0475 & {\ul 0.1040} & \textbf{0.1307} & 25.67\%    \\
    \multicolumn{1}{c|}{}                        & @10 & 0.0477  & 0.0407 & 0.0590 & 0.0613  & 0.0628 & 0.0694 & 0.0668 & 0.0698 & 0.0734 & {\ul 0.1441} & \textbf{0.1749} & 21.37\%    \\
    \multicolumn{1}{c|}{}                        & @20 & 0.0748  & 0.0592 & 0.0888 & 0.0942  & 0.0965 & 0.1065 & 0.0989 & 0.1058 & 0.1115 & {\ul 0.1928} & \textbf{0.2275} & 18.00\%    \\
    \multicolumn{1}{c|}{}                        & @50 & 0.1289  & 0.0897 & 0.1461 & 0.1555  & 0.1566 & 0.1751 & 0.1545 & 0.1712 & 0.1806 & {\ul 0.2764} & \textbf{0.3108} & 12.45\%    \\ \hline
    \multicolumn{1}{c|}{\multirow{4}{*}{MAP}}    & @5  & 0.0159  & 0.0147 & 0.0195 & 0.0213  & 0.0217 & 0.0238 & 0.0236 & 0.0234 & 0.0253 & {\ul 0.0640} & \textbf{0.0830} & 29.69\%    \\
    \multicolumn{1}{c|}{}                        & @10 & 0.0184  & 0.0166 & 0.0224 & 0.0242  & 0.0247 & 0.0271 & 0.0267 & 0.0268 & 0.0287 & {\ul 0.0693} & \textbf{0.0889} & 28.28\%    \\
    \multicolumn{1}{c|}{}                        & @20 & 0.0202  & 0.0179 & 0.0244 & 0.0265  & 0.0271 & 0.0296 & 0.0289 & 0.0293 & 0.0313 & {\ul 0.0727} & \textbf{0.0926} & 27.37\%    \\
    \multicolumn{1}{c|}{}                        & @50 & 0.0219  & 0.0188 & 0.0263 & 0.0284  & 0.0290 & 0.0318 & 0.0307 & 0.0314 & 0.0335 & {\ul 0.0754} & \textbf{0.0953} & 26.39\%    \\ \bottomrule
    \end{tabular}
    }

    \resizebox{0.85\linewidth}{!}{
        \begin{tabular}{cl|cccccc|cccc|cc}
            \toprule
                \multicolumn{2}{c|}{Models}                        & FREEDOM & SMORE  & GRCN   & LATTICE & BM3    & PGL          & SLMRec & LGMRec & MGCN         & DRAGON       & GTC             & Improved(\%) \\ \hline
                \multicolumn{1}{c|}{\multirow{4}{*}{NDCG}}   & @5  & 0.0267  & 0.0194 & 0.0220 & 0.0225  & 0.0219 & 0.0261       & 0.0234 & 0.0261 & 0.0263       & {\ul 0.0263} & \textbf{0.0283} & 7.60\%     \\
                \multicolumn{1}{c|}{}                        & @10 & 0.0335  & 0.0243 & 0.0286 & 0.0288  & 0.0287 & 0.0336       & 0.0296 & 0.0343 & 0.0337       & {\ul 0.0344} & \textbf{0.0356} & 3.49\%     \\
                \multicolumn{1}{c|}{}                        & @20 & 0.0412  & 0.0294 & 0.0363 & 0.0368  & 0.0374 & 0.0426       & 0.0361 & 0.0434 & 0.0432       & {\ul 0.0434} & \textbf{0.0442} & 1.84\%     \\
                \multicolumn{1}{c|}{}                        & @50 & 0.0517  & 0.0353 & 0.0483 & 0.0491  & 0.0503 & {\ul 0.0564} & 0.0460 & 0.0558 & 0.0530       & 0.0542       & \textbf{0.0576} & 2.13\%     \\ \hline
                \multicolumn{1}{c|}{\multirow{4}{*}{Recall}} & @5  & 0.0390  & 0.0296 & 0.0326 & 0.0342  & 0.0327 & 0.0392       & 0.0352 & 0.0399 & {\ul 0.0400} & 0.0394       & \textbf{0.0409} & 2.25\%     \\
                \multicolumn{1}{c|}{}                        & @10 & 0.0617  & 0.0449 & 0.0527 & 0.0531  & 0.0533 & 0.0618       & 0.0540 & 0.0644 & 0.0627       & {\ul 0.0657} & \textbf{0.0676} & 2.89\%     \\
                \multicolumn{1}{c|}{}                        & @20 & 0.0922  & 0.0647 & 0.0825 & 0.0840  & 0.0868 & {\ul 0.0970} & 0.0792 & 0.0917 & 0.0950       & 0.0960       & \textbf{0.1008} & 3.92\%     \\
                \multicolumn{1}{c|}{}                        & @50 & 0.1446  & 0.0940 & 0.1413 & 0.1451  & 0.1507 & {\ul 0.1653} & 0.1277 & 0.1629 & 0.1441       & 0.1501       & \textbf{0.1709} & 3.39\%    \\ \hline
                \multicolumn{1}{c|}{\multirow{4}{*}{MAP}}    & @5  & 0.0218  & 0.0159 & 0.0179 & 0.0182  & 0.0178 & 0.0211       & 0.0190 & 0.0182 & 0.0211       & {\ul 0.0211} & \textbf{0.0231} & 9.48\%     \\
                \multicolumn{1}{c|}{}                        & @10 & 0.0246  & 0.0179 & 0.0206 & 0.0207  & 0.0205 & 0.0241       & 0.0214 & 0.0206 & 0.0241       & {\ul 0.0243} & \textbf{0.0261} & 7.41\%     \\
                \multicolumn{1}{c|}{}                        & @20 & 0.0267  & 0.0193 & 0.0226 & 0.0228  & 0.0229 & {\ul 0.0265} & 0.0232 & 0.0224 & 0.0260       & 0.0263       & \textbf{0.0286} & 7.92\%     \\
                \multicolumn{1}{c|}{}                        & @50 & 0.0283  & 0.0202 & 0.0245 & 0.0248  & 0.0249 & {\ul 0.0283} & 0.0247 & 0.0239 & 0.0281       & 0.0284       & \textbf{0.0289} & 2.12\%    \\ \bottomrule
        \end{tabular}
    }

    \resizebox{0.85\linewidth}{!}{
        \begin{tabular}{cl|cccccc|cccc|cc}
            \toprule
                \multicolumn{2}{c|}{Models}                        & FREEDOM & SMORE        & GRCN   & LATTICE & BM3    & PGL    & SLMRec & LGMRec & MGCN         & DRAGON & GTC             & Improved(\%) \\ \hline
                \multicolumn{1}{c|}{\multirow{4}{*}{NDCG}}   & @5  & 0.0508  & {\ul 0.0531} & 0.0439 & 0.0477  & 0.0504 & 0.0508 & 0.0493 & 0.0515 & 0.0518       & 0.0499 & \textbf{0.0547} & 3.01\%     \\
                \multicolumn{1}{c|}{}                        & @10 & 0.0618  & {\ul 0.0650} & 0.0545 & 0.0587  & 0.0614 & 0.0622 & 0.0605 & 0.0636 & 0.0639       & 0.0629 & \textbf{0.0670} & 3.08\%     \\
                \multicolumn{1}{c|}{}                        & @20 & 0.0747  & {\ul 0.0776} & 0.0658 & 0.0706  & 0.0730 & 0.0744 & 0.0718 & 0.0764 & 0.0766       & 0.0755 & \textbf{0.0801} & 3.22\%     \\
                \multicolumn{1}{c|}{}                        & @50 & 0.0910  & {\ul 0.0950} & 0.0813 & 0.0864  & 0.0896 & 0.0927 & 0.0871 & 0.0942 & 0.0934       & 0.0926 & \textbf{0.0982} & 3.37\%     \\ \hline
                \multicolumn{1}{c|}{\multirow{4}{*}{Recall}} & @5  & 0.0760  & {\ul 0.0778} & 0.0666 & 0.0708  & 0.0747 & 0.0753 & 0.0744 & 0.0770 & 0.0778       & 0.0745 & \textbf{0.0796} & 2.31\%     \\
                \multicolumn{1}{c|}{}                        & @10 & 0.1101  & 0.1126       & 0.0992 & 0.1049  & 0.1085 & 0.1105 & 0.1090 & 0.1147 & {\ul 0.1150} & 0.1144 & \textbf{0.1185} & 3.04\%     \\
                \multicolumn{1}{c|}{}                        & @20 & 0.1604  & 0.1634       & 0.1436 & 0.1518  & 0.1542 & 0.1583 & 0.1532 & 0.1643 & {\ul 0.1648} & 0.1640 & \textbf{0.1693} & 2.73\%     \\
                \multicolumn{1}{c|}{}                        & @50 & 0.2423  & {\ul 0.2515} & 0.2208 & 0.2307  & 0.2374 & 0.2498 & 0.2297 & 0.2505 & 0.2485       & 0.2496 & \textbf{0.2572} & 2.27\%     \\ \hline
                \multicolumn{1}{c|}{\multirow{4}{*}{MAP}}    & @5  & 0.0421  & {\ul 0.0432} & 0.0360 & 0.0396  & 0.0421 & 0.0424 & 0.0406 & 0.0429 & 0.0429       & 0.0414 & \textbf{0.0442} & 2.31\%     \\
                \multicolumn{1}{c|}{}                        & @10 & 0.0466  & {\ul 0.0481} & 0.0403 & 0.0441  & 0.0466 & 0.0470 & 0.0451 & 0.0478 & 0.0478       & 0.0466 & \textbf{0.0491} & 2.08\%     \\
                \multicolumn{1}{c|}{}                        & @20 & 0.0501  & {\ul 0.0516} & 0.0433 & 0.0474  & 0.0497 & 0.0503 & 0.0482 & 0.0513 & 0.0512       & 0.0500 & \textbf{0.0545} & 5.62\%     \\
                \multicolumn{1}{c|}{}                        & @50 & 0.0527  & {\ul 0.0544} & 0.0458 & 0.0499  & 0.0523 & 0.0532 & 0.0506 & 0.0541 & 0.0539       & 0.0528 & \textbf{0.0563} & 3.49\%     \\ \bottomrule
        \end{tabular}
    }
\vspace{-2mm}
\end{table*}

\subsection{Computational Complexity}
The computational complexity of GTC mainly contains three components:
(1) {\tt Encoders}. GTC employs three parallel LightGCNs, resulting in $\mathcal{O}(L |\gE| d)$.
(2) {\tt Interaction-guided Denoising} and {\tt Content Features Generation}.
Interaction-guided Denoising samples a single timestep $t$, constructing noisy content representation $\rmC_t$ from $\rmC_0$ via the closed-form forward process.
Ignoring the negligible $\rmO(Bd)$ noise sampling and element-wise operations, the per-iteration time complexity is $\rmO(cost_{\phi}(B,d))$, where $cost_{\phi}(B,d)$ denotes the complexity of a single-forward pass of the U-Net network $\boldsymbol{\epsilon}_{\phi}$ on a mini-batch of size $B$ with $d$-dimensional content representations.
In the Content Features Generation stage, we iteratively apply the learned reverse diffusion from $\rmC_T$ to $\rmC_0$.
Each of the $T$ reverse steps requires one forward pass of $\boldsymbol{\epsilon}_\phi$, yielding overall complexity $\rmO(T \cdot cost_{\phi}(B,d))$.
(2) {\tt Total Correlation}. $\gL_{\rm CON}$ is implemented as a multi-modal InfoNCE loss over $N$ triplets.
We compute scores against all in-batch candidates for each anchor using multilinear function $h(\cdot)$.
This all-pairs construction yields $\rmO(N^2d)$.
By adding three such terms, which add only a constant factor, the overall complexity remains $\rmO(N^2d)$.

\section{Experiments}\label{sec:experiments}
In this section, we conduct a set of experiments to validate GTC, structured around six research questions (RQs):
\begin{itemize}
    \item[RQ1)] Can GTC outperform selected MMR baselines?
    \item[RQ2)] How does each component contribute to performance?
    \item[RQ3)] What are the impacts of visual and textual modalities?
    \item[RQ4)] Does GTC effectively ensure cross-modal balance?
    \item[RQ5)] Does GTC guarantee user preference consistency?
    \item[RQ6)] How do hyperparameters affect the GTC's performance?
\end{itemize}

\begin{table}
    \caption{Dataset statistics.}
    \vspace{-3mm}
    \label{tab: dataset}
    \centering
    \resizebox{\linewidth}{!}{
        \begin{tabular}{l|c|c|cc|cc}
            \toprule
            \multirow{2}{*}{Datasets} & \multirow{2}{*}{\# interaction} & \multirow{2}{*}{sparsity} & \multicolumn{2}{c|}{user} & \multicolumn{2}{c}{item}                            \\ \cline{4-7}
                                      &                                 &                           & num                       & avg interaction          & num    & avg interaction \\ \midrule
            Sports                    & 296,337                         & 99.95\%                   & 35,598                    & 8.3245                   & 18,357 & 16.1430         \\ \hline
            Baby                      & 160,792                         & 99.88\%                   & 19,445                    & 8.2691                   & 7,050  & 22.8074         \\ \hline
            Cell                      & 194,439                         & 99.93\%                   & 27,879                    & 6.9744                   & 10,429 & 18.6441         \\ \bottomrule
        \end{tabular}
    }
    \vspace{-5mm}
\end{table}

\subsection{Experimental Setup}
\paragraph{Datasets.}
Following standard practice~\cite{zhou2023bootstrap}, we evaluate GTC on 3 widely used public benchmarks from the Amazon Review Datasets: {\bf Sports and Outdoors} (Sports), {\bf Baby} (Baby), and {\bf Cellphone} (Cell).
We filter for users and items with at least five interactions and randomly split the data into $80\%$ for training, $10\%$ for validation, and $10\%$ for testing.
The dataset statistics are provided in Table. \ref{tab: dataset}.


\paragraph{Baselines.}
We compare GTC against 10 recent MMR methods, including fusion-based methods (FREEDOM~\cite{zhou2023tale}, GRCN~\cite{wei2020graph}, SMORE~\cite{SMORE}, LATTICE~\cite{zhang2021mining}, MGCN~\cite{yu2023multi}, PGL~\cite{yu2025mind}) and distenglement-based methods (SLMRec~\cite{tao2022self}, DRAGON~\cite{zhou2023enhancing}, BM3~\cite{zhou2023bootstrap}, LGMRec~\cite{guo2024lgmrec}).

\paragraph{Implementation.}
For content modality encoders, ResNet-50~\citep{he2016deep} and sentence-transformer~\citep{reimers2019sentence} are used to extract visual features ($d_v=4096$) and textual features ($d_t=384$).
All baselines and GTC are trained using the Adam optimizer with a learning rate of $0.001$ and a representation dimension of $64$.
For GTC, the diffusion process uses $T=500$ timesteps for Sports, $T=600$ for Baby, and $T=700$ for Cell.
The regularization weight is set to $0.01$.
In the interaction-guided denoising, the noise starts at \(\beta_s=1e-4\) and increases to \(\beta_t=0.02\).
Experiments show that variations in the noise levels only affect the model convergence speed and have no obvious impact on model performance.
The weights $w_1$ and $w_2$ are searched in $\{0.1, 0.2, 0.3, 0.4, 0.5, 0.6, 0.7, 0.8, 0.9, 1\}$.
For noise steps, we search in \{100, 200, 300, 400, 500, 600, 700, 800, 900, 1000\}.
We report recommendation performance using Normalized Discounted Cumulative Gain~(NDCG@K), Mean Average Precision~(MAP@K), and F1@K for top-$K$ items with $K \in \{5, 10, 20, 50 \}$.
We run experiments with 10 different random seeds and report the final results.
All experiments are run on either a single NVIDIA V100, DGX A100, or NVIDIA RTX A5000 GPU.

\subsection{Overall Performance (RQ1)}
Table.~\ref{tab: sportsresult} shows the main results.
GTC consistently and significantly outperforms all baselines across all datasets and metrics.
On the Sports dataset, GTC achieves $28.30\%$ improvement in NDCG@5 over the strongest baseline.
The performance gains on the Baby dataset are also consistent, reaching up to $9.48\%$ in MAP@5.
These results provide an affirmative answer to RQ1.
Moreover, we observe the following patterns reflecting the limitations of prior methods.

\paragraph{Inconsistent Multi-Modal Features Hinder Simple Fusion.}
SMORE and GRCN generate features for each modality in isolation and combine them via simple fusion, leading to the weakest performance.
LGMRec and PGL explore multimodal information by capturing global and local structural patterns, but they still lack mechanisms to remove irregular noise.
FREEDOM and LATTICE attempt to harness textual and visual features in modeling user-item interactions, but they merely use learnable weights to differentiate the modality importance and aggregate them without considering their mutual entanglement.
This suggests that overlooking the inconsistencies and potential noise between modalities limits the quality of MMR.

\paragraph{Refinement is Beneficial.}
Methods like BM3, SLMRec, MGCN, and DRAGON, which incorporate modules to denoise or align multi-modal features, demonstrate stronger performance relative to fusion-based methods.
Specifically, DRAGON's empirical evaluation shows that indiscriminately fusing multi-modal features can reduce overall performance \cite{zhou2023enhancing}.
MGCN argues that conventional fusion techniques, such as concatenation or mean-pooling, fail to capture the ever-changing importance of features, and therefore implements an attention network to adjust the importance of each modality dynamically.
BM3 and SLMRec adopt contrastive learning to ensure effective cross-modal alignment.
Their performances confirm the importance of addressing modality inconsistency.

\vspace{-1mm}
\paragraph{Pairwise Alignment is Still a Bottleneck.}
However, these top-performing baselines rely on aligning modalities through pairwise contrastive objectives.
To generate refined embeddings for feature fusion, SLMRec employs DRL-based techniques to build hypergraphs that capture both shared and unique modality features by optimizing contrastive loss.
Likewise, BM3 and MGCN perform inter-modality alignment by focusing on pairwise correlations through contrastive learning.
GTC, however, leads over even the strongest baseline, providing evidence that this pairwise approach is a performance bottleneck.
It supports our claim that capturing the higher-order correlations unlocks a new level of performance.

\vspace{-3mm}
\subsection{Ablation Study (RQ2)}

\begin{table}[]
    \caption{Ablation experiments on Sports (up), Baby dataset (middle), and Cell (bottom).}
    \label{tab: ablation}
    \centering
    \vspace{-3mm}
    \resizebox{\linewidth}{!}{
        \begin{tabular}{l|cccc|cccc}
        \toprule
        \multicolumn{1}{c|}{\multirow{2}{*}{Module}} & \multicolumn{4}{c|}{Recall}       & \multicolumn{4}{c}{NDCG}          \\ \cline{2-9} 
        \multicolumn{1}{c|}{}                        & @5     & @10    & @20    & @50    & @5     & @10    & @20    & @50    \\ \hline
        GTC Base                & 0.0995                 & 0.1367                  & 0.1841                  & 0.2610                  & 0.0709                 & 0.0831                  & 0.0953                  & 0.1109                  \\
        GTC Base w/ DN          & {\ul 0.1229}           & {\ul 0.1663}            & {\ul 0.2180}            & {\ul 0.3020}            & {\ul 0.0894}           & {\ul 0.1037}            & {\ul 0.1170}            & {\ul 0.1340}            \\
        GTC Base w/ TC          & 0.1206                 & 0.1641                  & 0.2164                  & 0.2994                  & 0.0884                 & 0.1027                  & 0.1162                  & 0.1330                  \\
        GTC w/o TC              & 0.1099                 & 0.1506                  & 0.1982                  & 0.2794                  & 0.0785                 & 0.0918                  & 0.1041                  & 0.1206                  \\ \hline
        GTC                     & \textbf{0.1307}        & \textbf{0.1749}         & \textbf{0.2275}         & \textbf{0.3108}         & \textbf{0.0961}        & \textbf{0.1106}         & \textbf{0.1242}         & \textbf{0.1411}        \\ \bottomrule       
    \end{tabular}}

    \resizebox{\linewidth}{!}{
        \begin{tabular}{l|cccc|cccc}
        \toprule
        \multicolumn{1}{c|}{\multirow{2}{*}{Module}} & \multicolumn{4}{c|}{Recall}       & \multicolumn{4}{c}{NDCG}          \\ \cline{2-9} 
        \multicolumn{1}{c|}{}                        & @5     & @10    & @20    & @50    & @5     & @10    & @20    & @50    \\ \hline
        GTC Base                & 0.0391                 & 0.0619                  & 0.0976                  & 0.1667                  & 0.0263                 & 0.0337                  & 0.0429                  & 0.0569                  \\
        GTC Base w/ DN          & {\ul 0.0405}           & {\ul 0.0668}            & {\ul 0.1004}            & {\ul 0.1699}            & {\ul 0.0278}           & {\ul 0.0354}            & {\ul 0.0433}            & {\ul 0.0582}            \\
        GTC Base w/ TC          & 0.0402                 & 0.0660                  & 0.1001                  & 0.1687                  & 0.0268                 & 0.0352                  & 0.0431                  & 0.0579                  \\
        GTC w/o TC              & 0.0401                 & 0.0652                  & 0.0996                  & 0.1686                  & 0.0272                 & 0.0353                  & 0.0430                  & 0.0576                  \\
        GTC                     & \textbf{0.0409}        & \textbf{0.0676}         & \textbf{0.1008}         & \textbf{0.1709}         & \textbf{0.0283}        & \textbf{0.0356}         & \textbf{0.0442}         & \textbf{0.0585}   \\ \bottomrule        
    \end{tabular}}

    \resizebox{\linewidth}{!}{
        \begin{tabular}{l|cccc|cccc}
        \toprule
        \multicolumn{1}{c|}{\multirow{2}{*}{Module}} & \multicolumn{4}{c|}{Recall}       & \multicolumn{4}{c}{NDCG}          \\ \cline{2-9} 
        \multicolumn{1}{c|}{}                        & @5     & @10    & @20    & @50    & @5     & @10    & @20    & @50    \\ \hline
        GTC Base                                     & 0.0772          & 0.1126          & 0.1609          & 0.2441          & 0.0520          & 0.0635          & 0.0758          & 0.0924          \\
        GTC Base w/ DN                               & 0.0767          & 0.1141          & 0.1644          & 0.2459          & 0.0516          & 0.0638          & 0.0765          & 0.0929          \\
        GTC Base w/ TC                               & {\ul 0.0779}    & {\ul 0.1160}    & {\ul 0.1669}    & {\ul 0.2550}    & 0.0525          & 0.0648          & {\ul 0.0778}    & {\ul 0.0954}    \\
        GTC w/o TC                                   & 0.0777          & 0.1147          & 0.1640          & 0.2509          & {\ul 0.0527}    & {\ul 0.0648}    & 0.0773          & 0.0947          \\ \hline
        GTC                                          & \textbf{0.0796} & \textbf{0.1185} & \textbf{0.1693} & \textbf{0.2572} & \textbf{0.0547} & \textbf{0.0670} & \textbf{0.0801} & \textbf{0.0982}  \\ \bottomrule        
    \end{tabular}}
\vspace{-5mm}
\end{table}

To dissect the contribution of GTC's key components, we evaluate several variants on the three datasets, shown in Table.~\ref{tab: ablation}.
Namely, {\tt GTC Base} is minimal baseline using only LightGCNs with concatenation.
    {\tt GTC Base w/ DN} is the baseline equipped with the denoising process.
    {\tt GTC Base w/ TC} uses total correlation optimization (i.e., GTC variant removing denoising $\gL_{\rm GEN}$).
    {\tt GTC w/o TC} replaces the total correlation objective $\gL_{\rm CON}$ with a standard pairwise contrastive loss, while retaining the diffusion module.
The results clearly show the necessity of both components.
Interaction-guided denoising ({\tt GTC Base w/ DN}) yields a substantial performance improvement (e.g., an 2.34\% relative increase in Recall@5 in Sports) compared to {\tt GTC Base}, indicating that denoising content features with user-specific signals is critical.
Similarly, both total correlation ({\tt GTC Base w/ TC}) and pairwise contrastive objectives ({\tt GTC w/o TC}) provide consistent gains over {\tt GTC Base}, highlighting the benefit of enforcing cross-modal consistency.
However, {\tt GTC w/o TC} degrades {\tt GTC Base w/ TC}'s performance, supporting that higher-order dependencies are crucial for alignment.
The full GTC model, integrating both components, achieves the best performance.


\begin{figure}
    \centering
    \vspace{-5mm}
    \begin{subfigure}[b]{\linewidth}
        \centering
        \includegraphics[width=\textwidth]{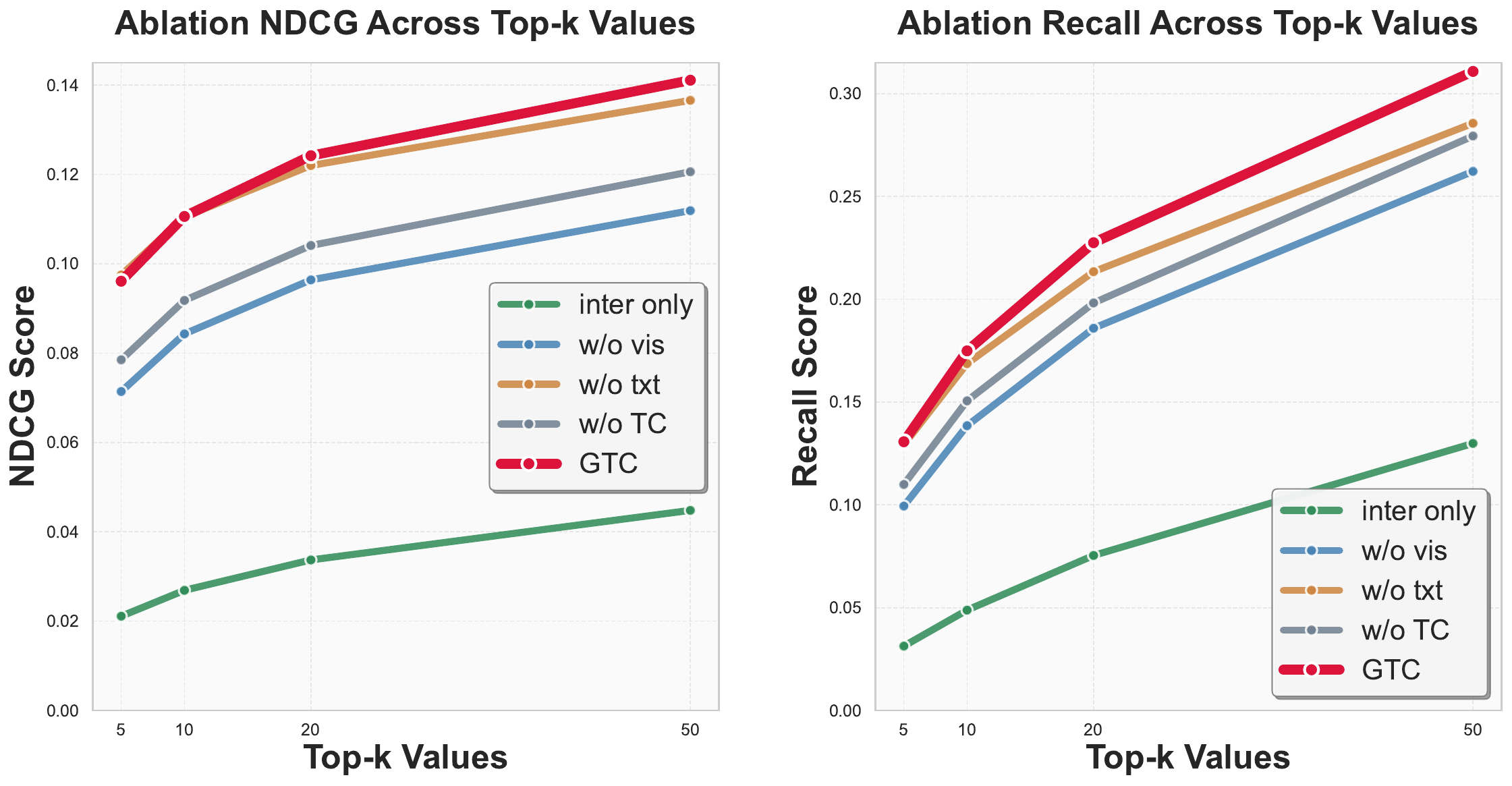}
        \label{fig:ds}
    \end{subfigure}
    \vspace{-4mm}
    \begin{subfigure}[b]{\linewidth}
        \centering
        \includegraphics[width=\textwidth]{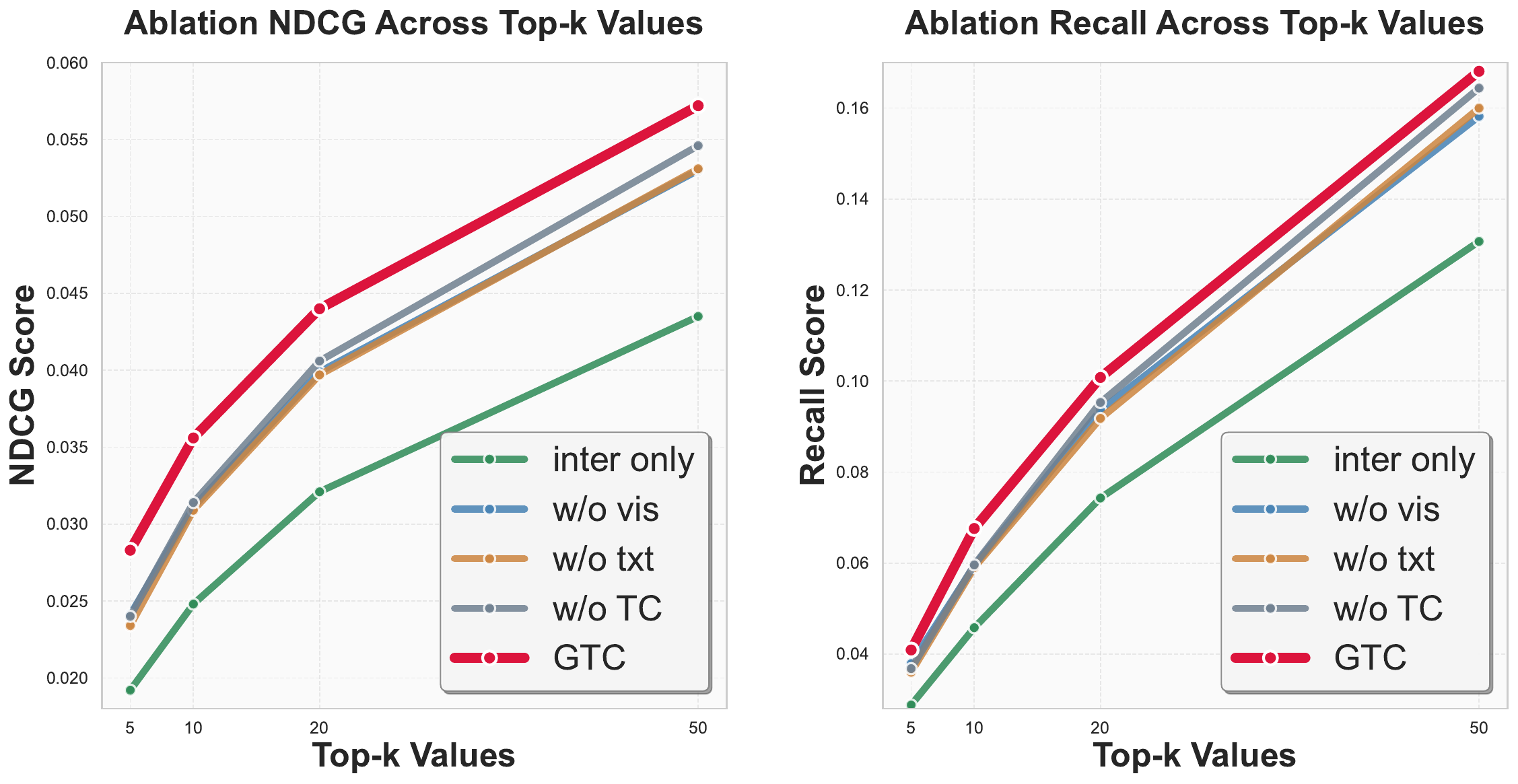}
        \label{fig:ds}
    \end{subfigure}
    \vspace{-4mm}
    \begin{subfigure}[b]{\linewidth}
        \centering
        \includegraphics[width=\textwidth]{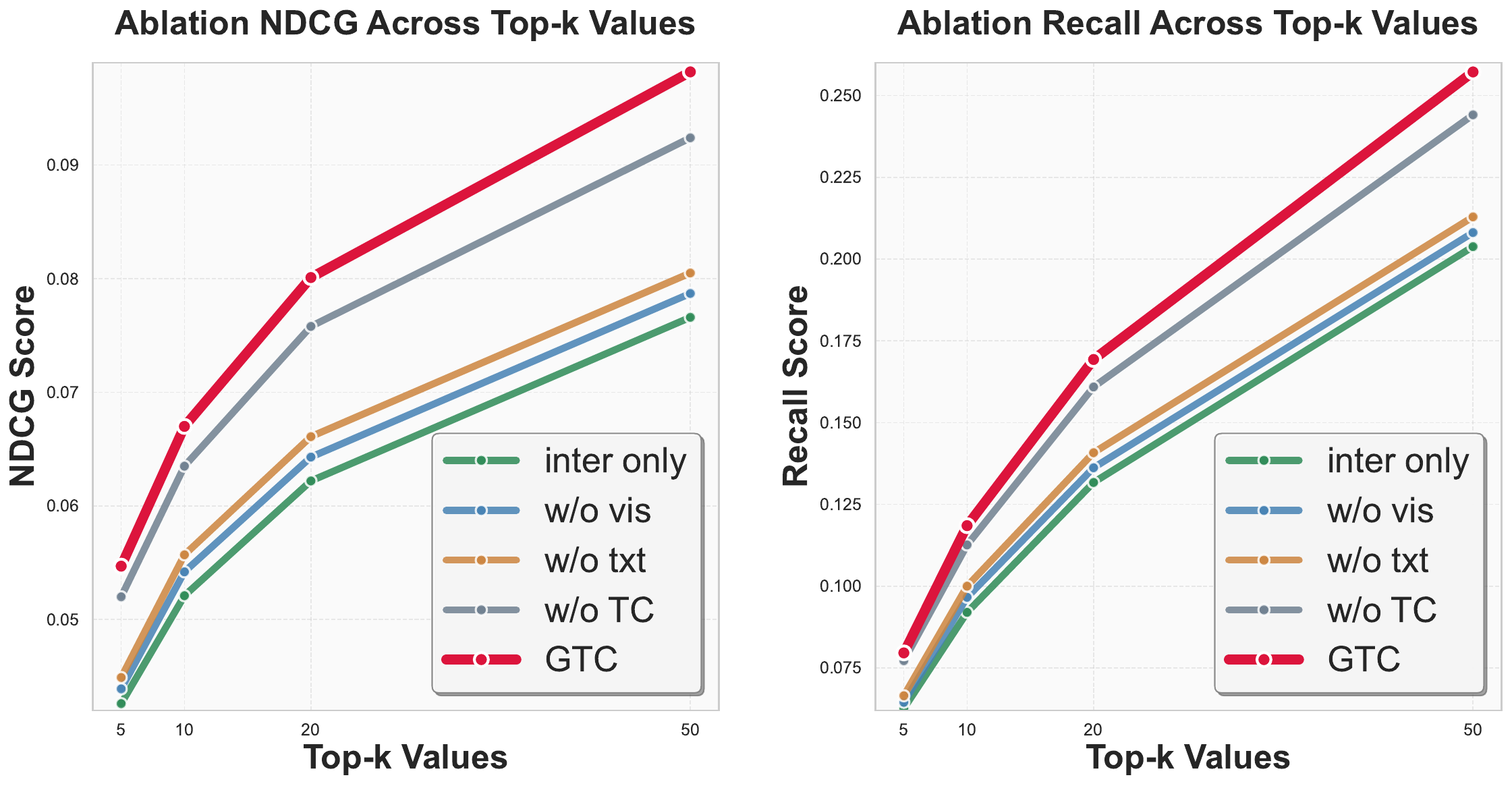}
        \label{fig:gs}
    \end{subfigure}
    \vspace{-4mm}
    \caption{Impact of content features in Sports (up), Baby (middle), and Cell (bottom) Datasets.}
    \label{fig: features}
    \vspace{-5mm}
\end{figure}

\subsection{Impact of Content Features (RQ3)}

Fig.~\ref{fig: features} shows the performance when modalities are selectively removed.
The {\tt inter-only} is a non-multimodal baseline that uses only user-item interaction data.
The {\tt w/o visual} and {\tt w/o textual} settings replace the total correlation approach with pairwise correlations implemented through standard contrastive learning as only two modalities are involved, confirming the value of side information with consistently improved performances over the {\tt inter-only} baseline.
The {\tt w/o TC} uses all three modalities but with pairwise alignment, which outperforms the two-modality versions, showing that more modalities provide richer information.
Still, GTC consistently outperforms all these variants on both Sports and Baby datasets.
This shows GTC's ability to not just use, but holistically integrate and refine information from all available modalities, unlocking performance unattainable by pairwise methods.

\subsection{Cross-modal Consistency (RQ4)}
Lastly, to assess whether GTC produces balanced and consistent cross-modal representations, we introduce the {\em modality balance score}.
For each user in the test set, we compute the cosine similarities ${\rm sim}(\bar{\rmS}, \bar{\rmE}_m)$ between the final fused representation $\bar{\rmS}$ and each of the three modality-specific representations $\bar{\rmE}_m$ for all $m \in \{\rmI, \rmV, \rmT \}$.
Specifically, this balance score is defined as $1 - (\max({\rm sim}(\bar{\rmS}, \bar{\rmE}_m)) - \min({\rm sim}(\bar{\rmS}, \bar{\rmE}_m)))$, where a higher score (closer to 1) indicates less disparity and thus a more balanced integration.

\begin{figure}
    \centering
    \includegraphics[width=\linewidth]{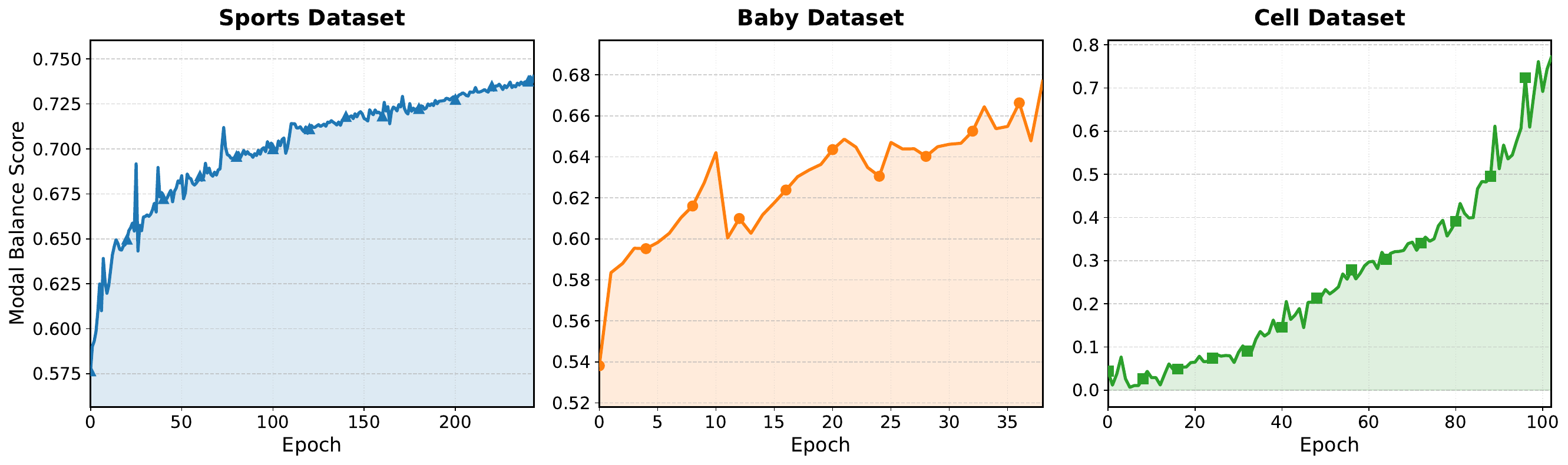}
    \vspace{-5mm}
    \caption{Modality balance trend during training GTC.}
    \label{fig:similarity}
    \vspace{-5mm}
\end{figure}

It is obvious from Fig.~\ref{fig:similarity} that, the {\em modality balance score} obtained by GTC steadily increases during training for both datasets as epochs increases.
This upward trend provides evidence that the GTC framework successfully learns to pull the representations from different modalities into a coherent space.
Notably, the curve for the Baby dataset is less smooth, suggesting that user preferences in this domain are more heterogeneous, making the alignment task more challenging.
Nonetheless, GTC is still able to handle effectively.

\begin{figure*}
    \centering
    \begin{subfigure}[b]{.33\linewidth}
        \centering
        \includegraphics[width=\textwidth]{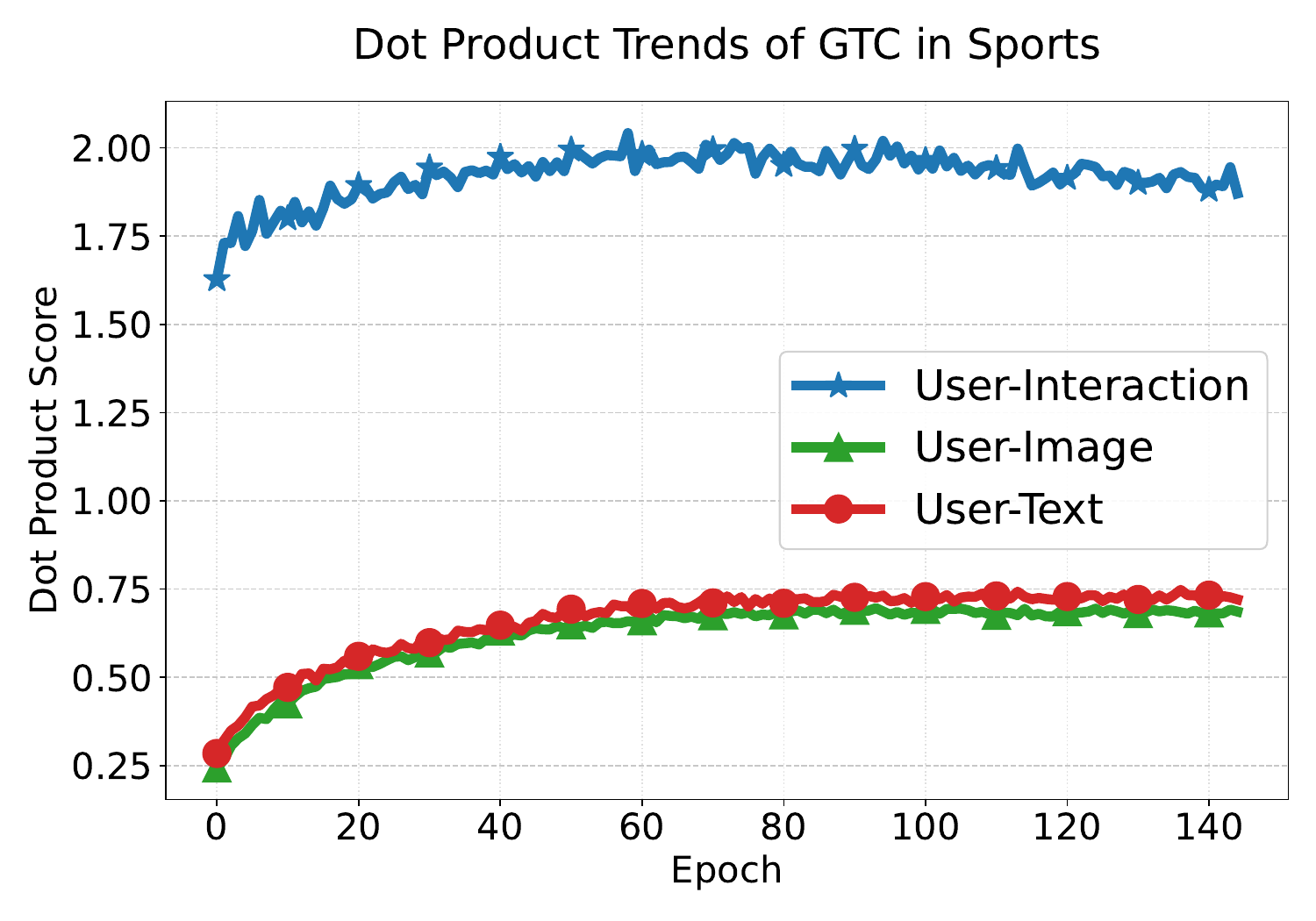}
        \label{fig:ds}
    \end{subfigure}
    \vspace{-3mm}
    \begin{subfigure}[b]{.33\linewidth}
        \centering
        \includegraphics[width=\textwidth]{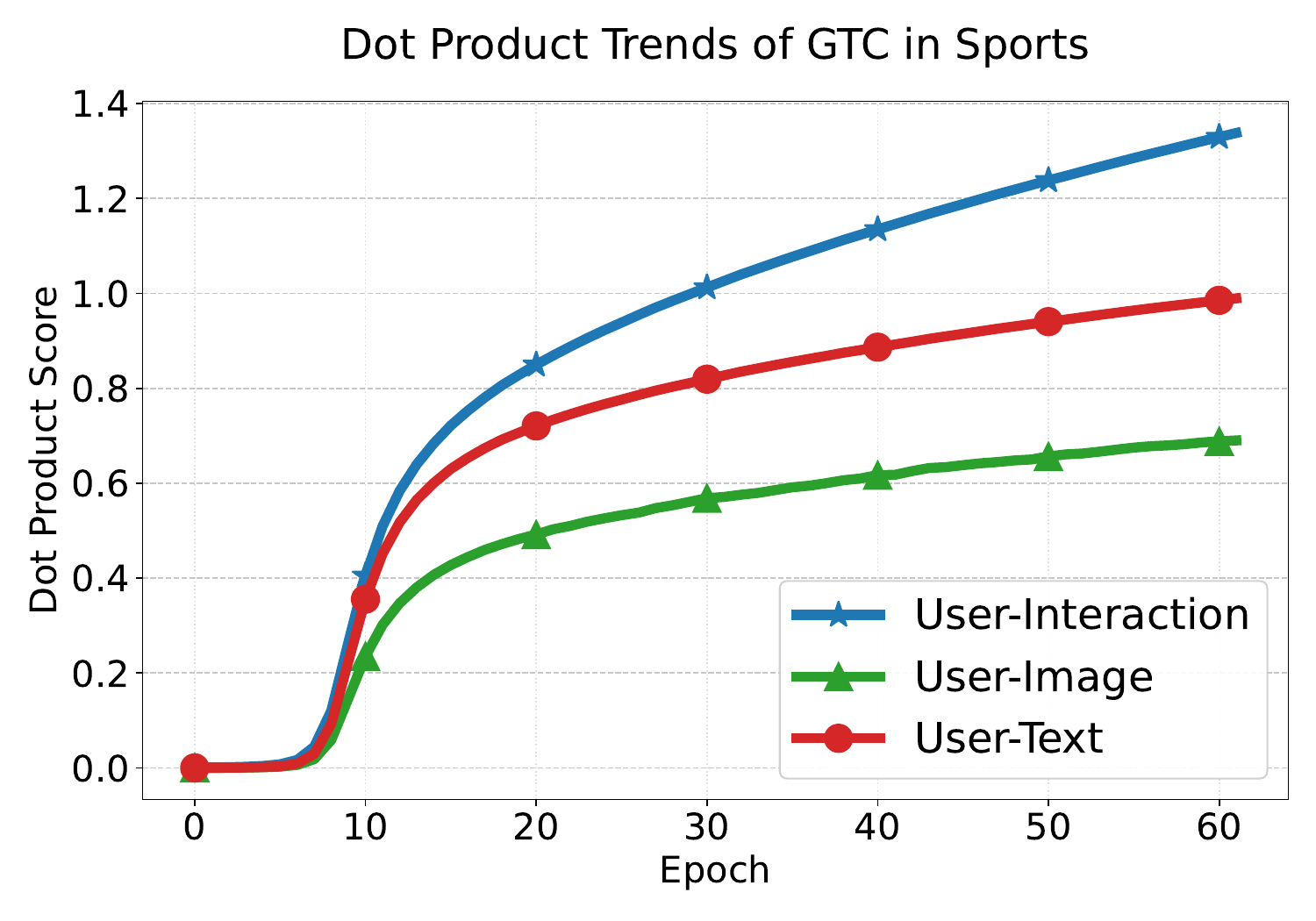}
        \label{fig:ds}
    \end{subfigure}
    \vspace{-3mm}
    \begin{subfigure}[b]{.33\linewidth}
        \centering
        \includegraphics[width=\textwidth]{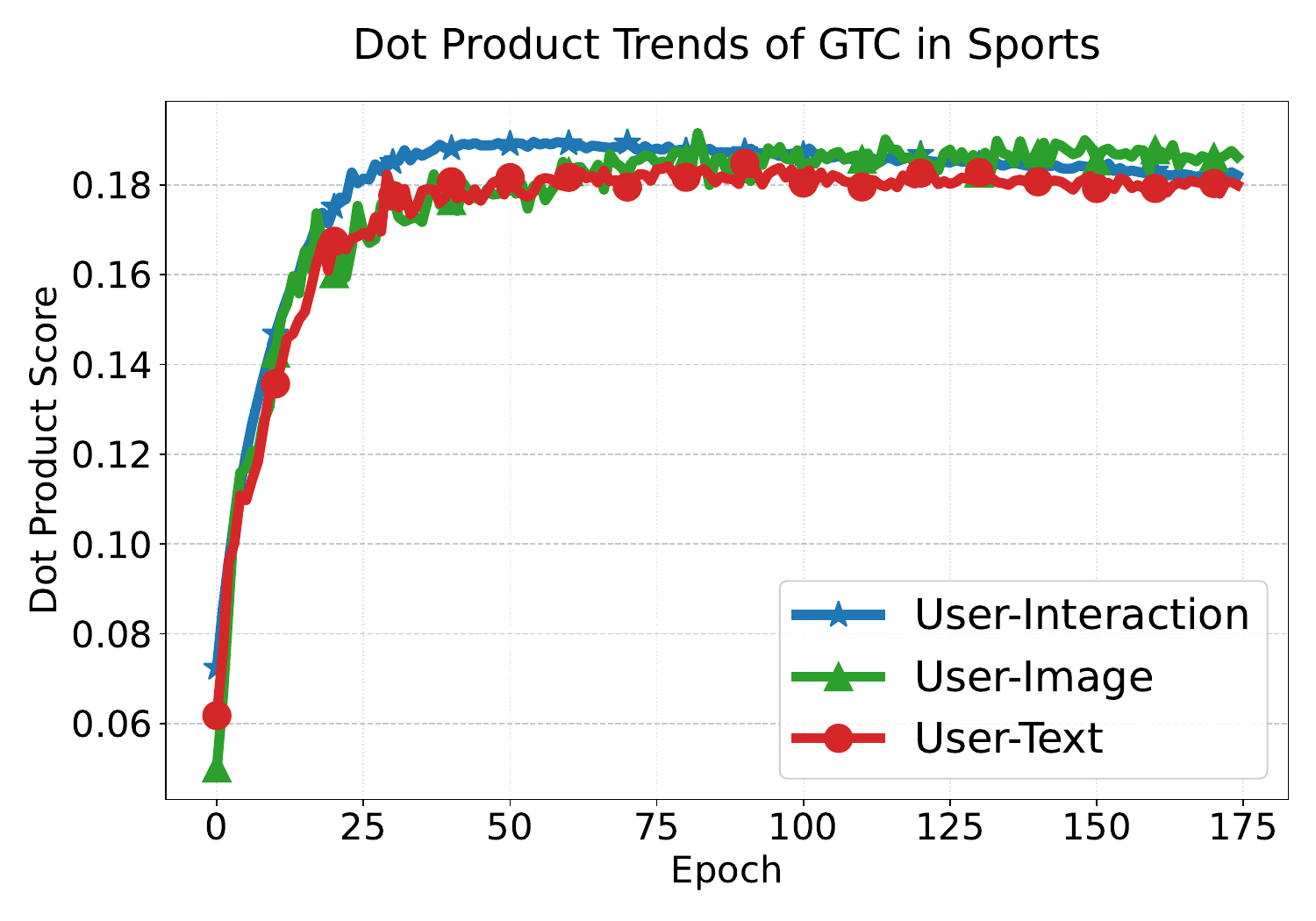}
        \label{fig:gs}
    \end{subfigure}
    \vspace{-3mm}

    \begin{subfigure}[b]{.33\linewidth}
        \centering
        \includegraphics[width=\textwidth]{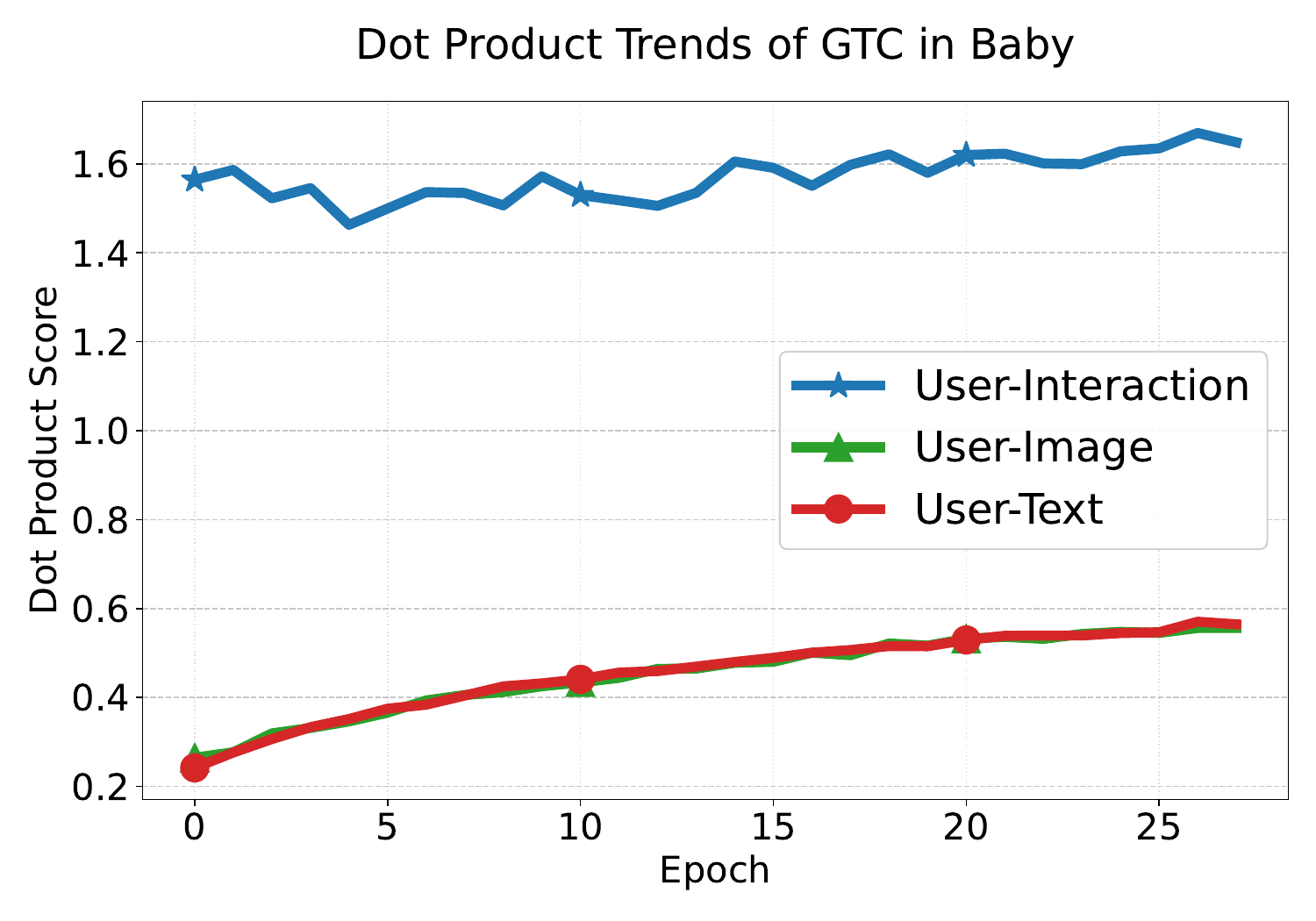}
        \label{fig:db}
    \end{subfigure}
    \vspace{-3mm}
    \begin{subfigure}[b]{.33\linewidth}
        \centering
        \includegraphics[width=\textwidth]{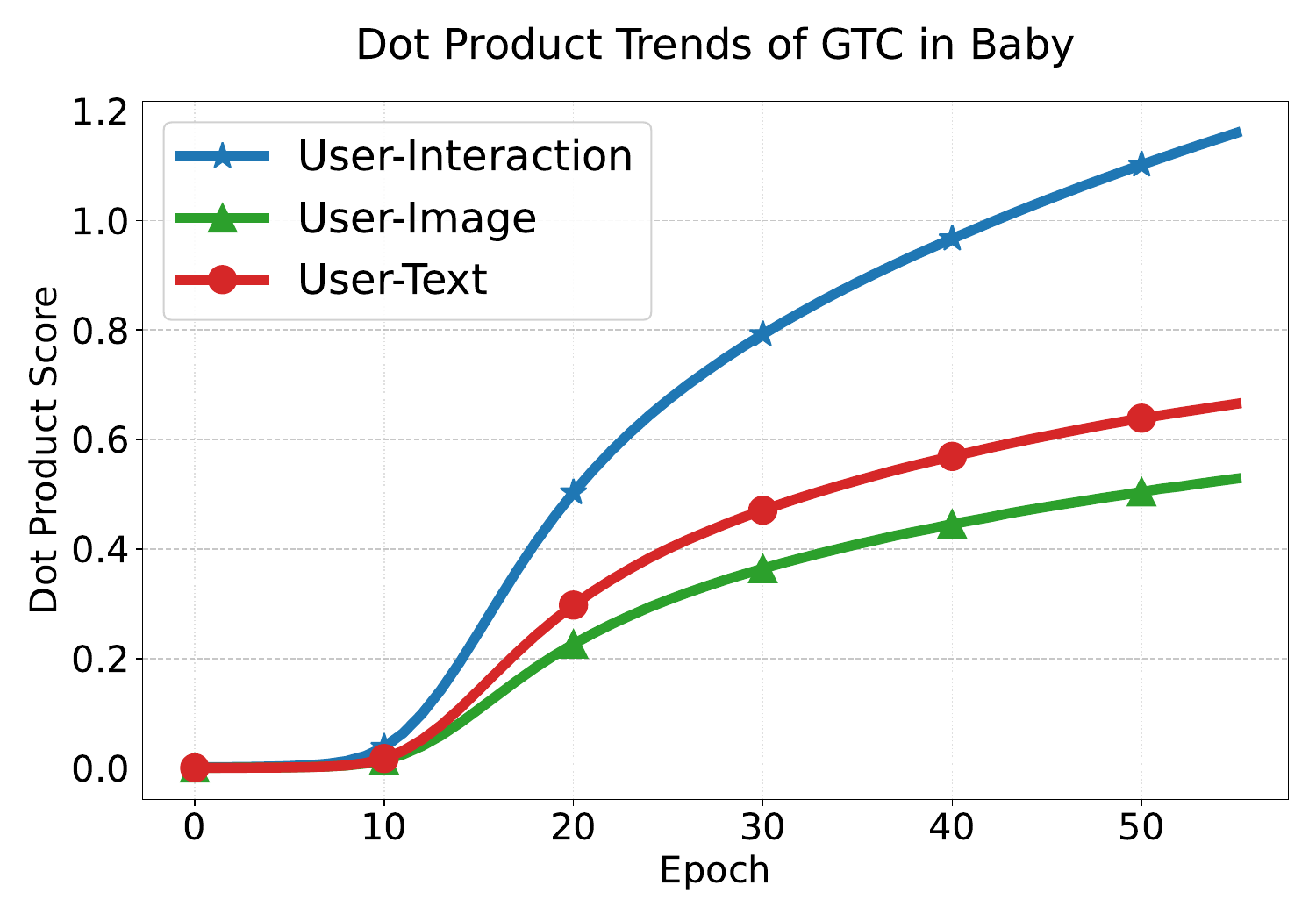}
        \label{fig:db}
    \end{subfigure}
    \vspace{-3mm}
    \begin{subfigure}[b]{.33\linewidth}
        \centering
        \includegraphics[width=\textwidth]{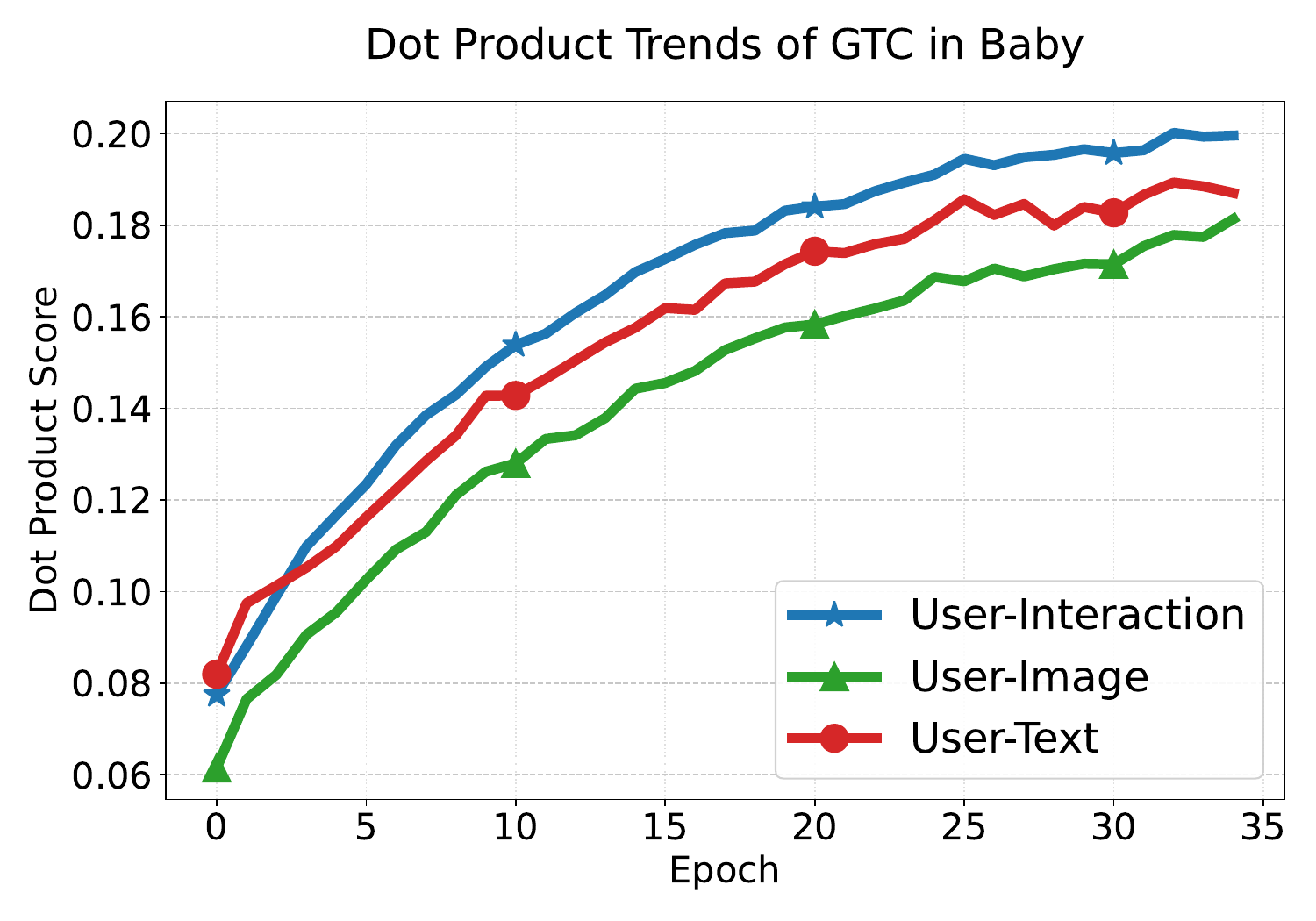}
        \label{fig:gb}
    \end{subfigure}
    \vspace{-3mm}
    
    \begin{subfigure}[b]{.33\linewidth}
        \centering
        \includegraphics[width=\textwidth]{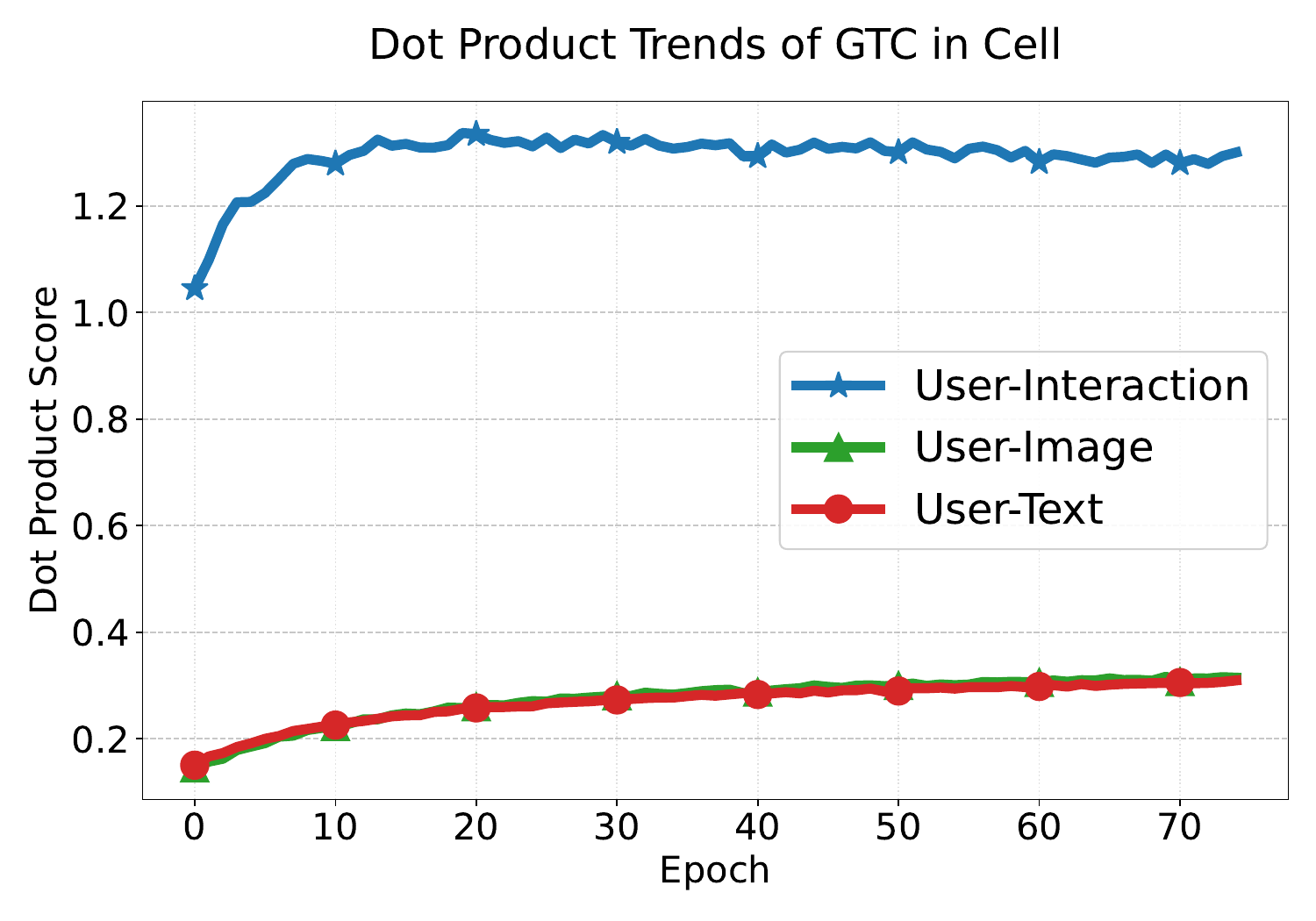}
        \label{fig:db}
    \end{subfigure}
    \vspace{-3mm}
    \begin{subfigure}[b]{.33\linewidth}
        \centering
        \includegraphics[width=\textwidth]{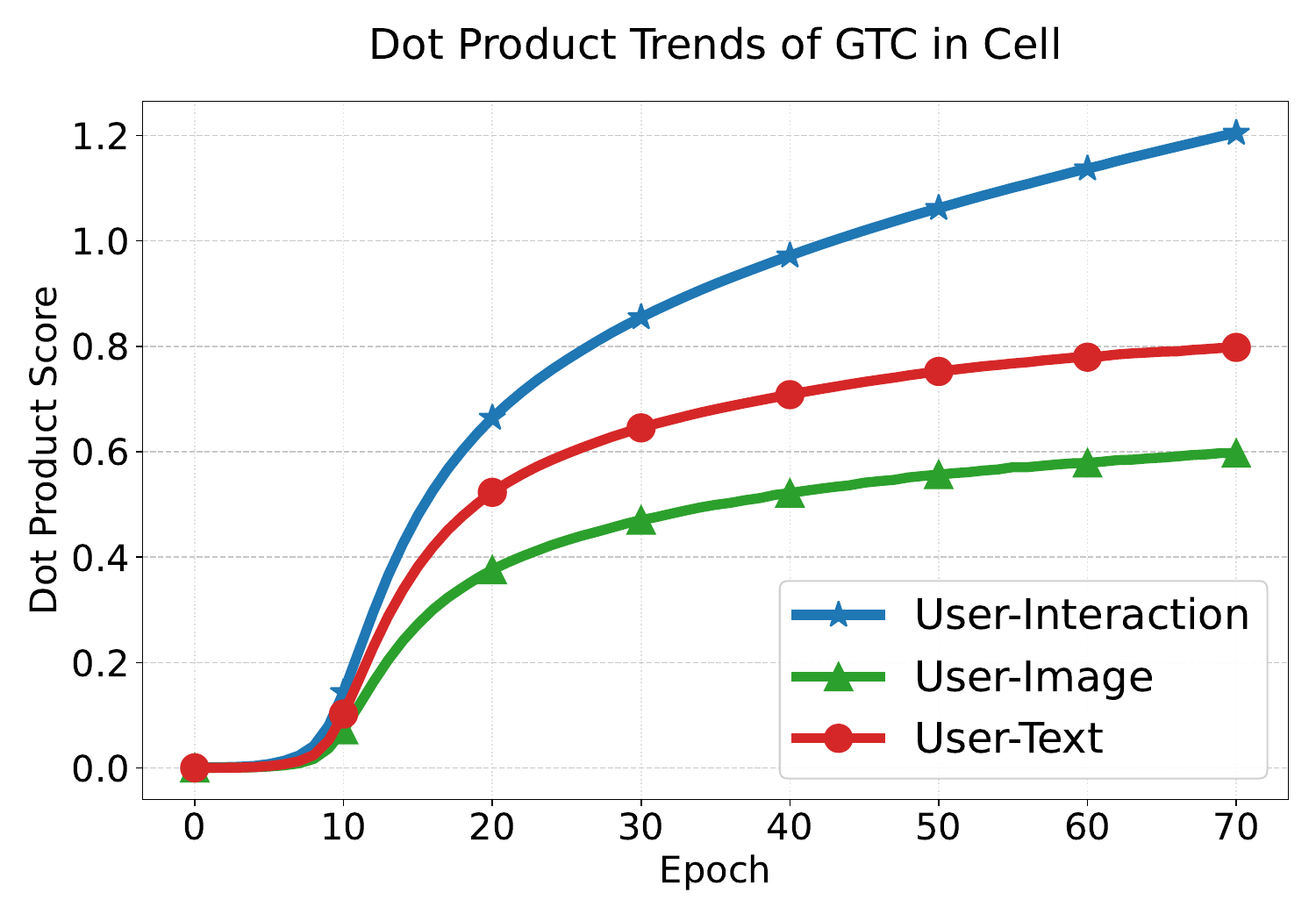}
        \label{fig:db}
    \end{subfigure}
    \vspace{-3mm}
    \begin{subfigure}[b]{.33\linewidth}
        \centering
        \includegraphics[width=\textwidth]{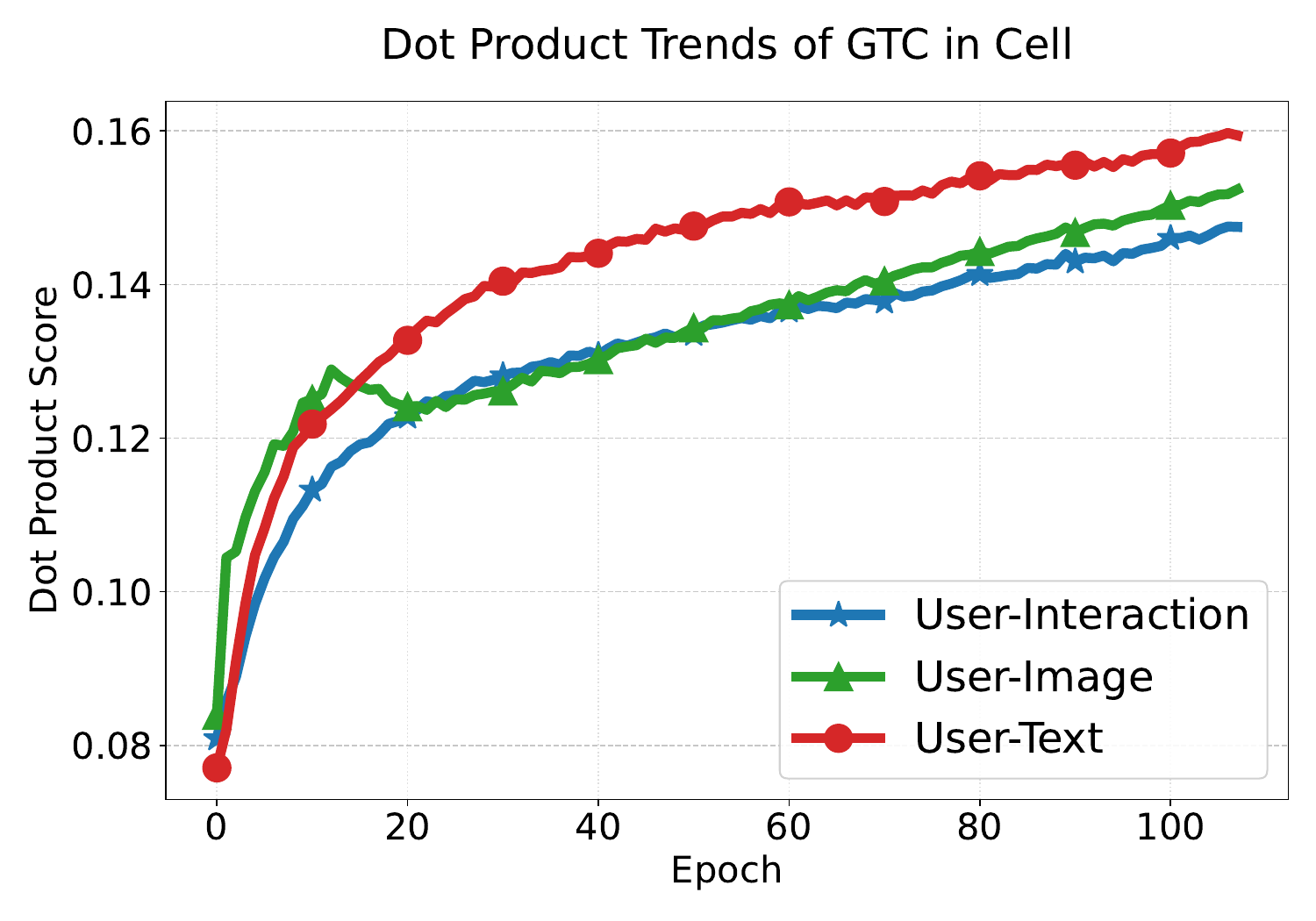}
        \label{fig:gb}
    \end{subfigure}
    \vspace{-3mm}
    \caption{User preference consistency in the Sports dataset (up) and Baby dataset (down).}
    \label{fig: consistency}
    \vspace{-3mm}

\end{figure*}

\subsection{User Preference Consistency (RQ5)}
To empirically validate the consistency achieved between learned modality features and user preferences, we visualize the dot product trends between user embeddings and the modality embeddings (i.e., $\bar{\rmS}, \bar{\rmV}, \bar{\rmT}$) within our GTC, the runner-up DRAGON, and MGCN (pairwise contrastive loss to enforce consistency). 
Crucially, we clarify that \textbf{a higher absolute dot product value does not inherently correlate with superior recommendation performance}. 

As shwon in Fig.~\ref{fig: consistency}, even without explicit constraints in DRAGON, dot products among three modalities gradually increase as the training process progresses, indicating that better performances implicitly require higher consistency between different modalities.
DRAGON's user-interaction curve consistently surpasses its user-image and user-text curves, suggesting a strong reliance on interaction signals with visual and textual modalities contributing to a more limited extent. 
Similarly, while MGCN utilizes pairwise contrastive loss to ensure steady upward trends of all modalities, it still suffers from an imbalance problem in modality contributions, especially late in training.
GTC shows \textit{a balanced upward trend across user-interaction, user-visual, and user-text curves}, signifying that its interaction-guided denoising mechanism effectively enhances the alignment of diverse modalities with user preference signals. 
This enhanced alignment leads to significant improvements in ranking results, confirming GTC's superior ability to leverage multimodal information for recommendations.

\begin{figure*}
    \centering
    \vspace{-3mm}
    \begin{subfigure}[b]{0.95\linewidth}
        \centering
        \includegraphics[width=\textwidth]{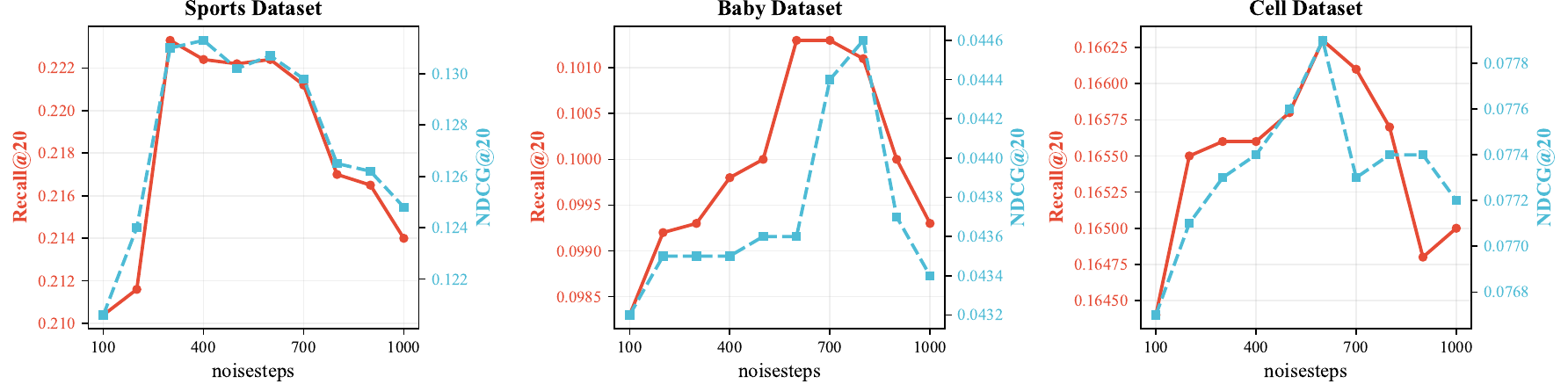}
        \label{fig:t_sports}
    \end{subfigure}
    \vspace{-3mm}
    \begin{subfigure}[b]{0.95\linewidth}
        \centering
        \includegraphics[width=\textwidth]{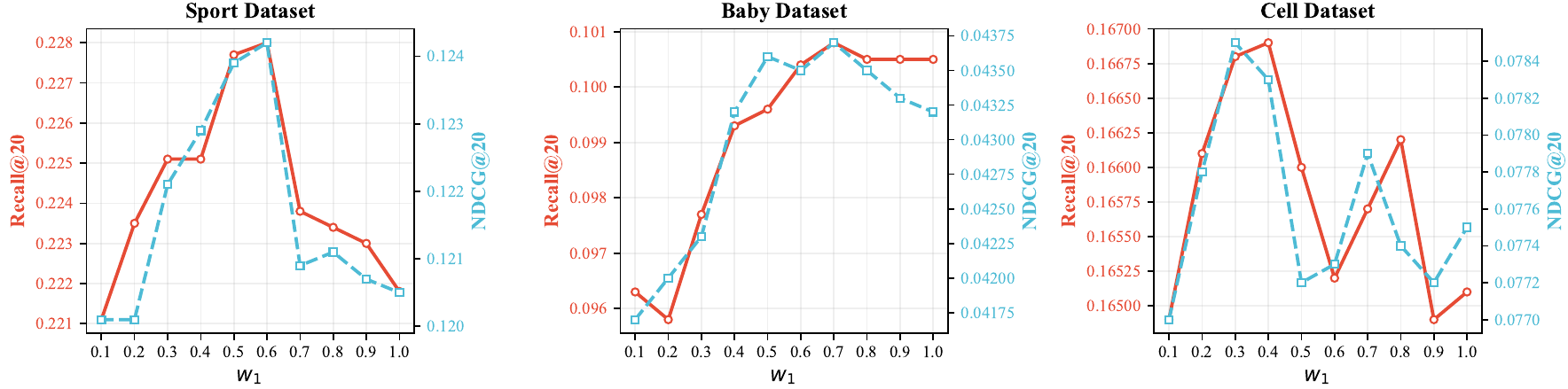}
        \label{fig:w1_sports}
    \end{subfigure}
    \vspace{-3mm}
    \begin{subfigure}[b]{0.95\linewidth}
        \centering
        \includegraphics[width=\textwidth]{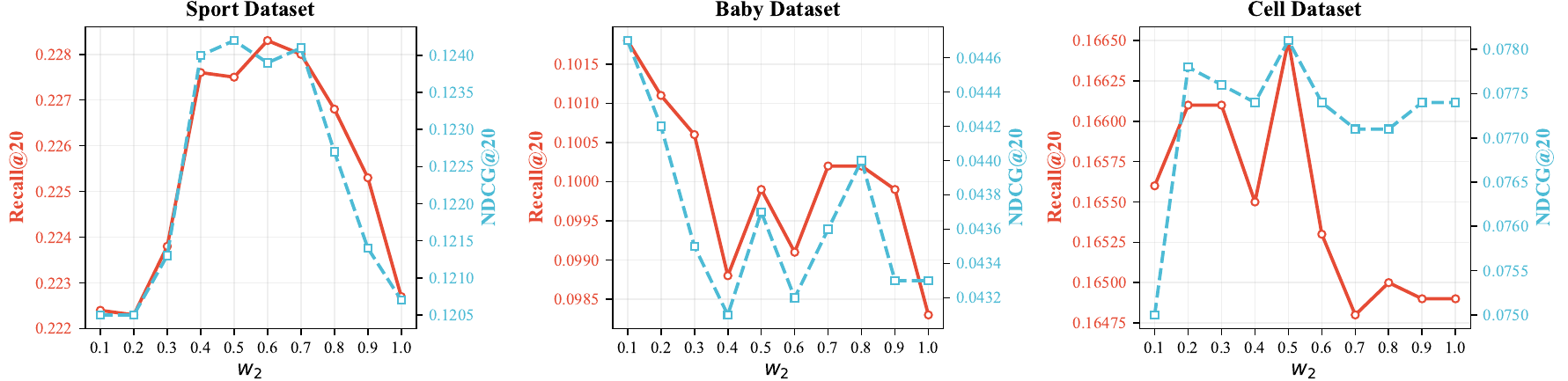}
        \label{fig:w2_sports}
    \end{subfigure}
    \vspace{-3mm}
    \caption{Parameter evaluation in Sports (left), Baby (middle), and Cell (right) datasets.}
    \vspace{-3mm}
    \label{fig: parameters}

\end{figure*}

\vspace{-3mm}
\subsection{Hyperparameter Evaluation (RQ6)}
GTC involves 3 important hyperparameters: noise step $T$, diffusion weight $w_1$, and total correlation weight $w_2$.
To investigate the best configuration, we conduct the hyperparameter search on three datasets, shown in Fig.~\ref{fig: parameters}.

In the \textit{user-aware content feature generation}, noise step $T$ controls the granularity of the noise injection and subsequent removal.
A larger $T$ implies a more gradual denoising, while a smaller $T$ enforces a coarse, abrupt reversal of noise.
On the Sports dataset, the best performance is obtained at $T$=300.
Differently, on Baby and Cell, GTC performs best with larger $T$, achieving the highest scores at $T$=600 and $T$=700, respectively, indicating that these datasets benefit from finer-grained discrimination than Sports.
Given the different scales and sparsity levels of user–item interactions, this pattern suggests that \textit{when interactions are relatively scarce or highly sparse (as in Baby and Cell), a more fine-grained denoising schedule enables the model to better exploit cross-modal consistency}, leading to improved performance.
In all three datasets, performance rises as $T$ increases, and subsequently declines.
This trend confirms that \textit{moderate noise injection is beneficial for aligning modality-specific signals, whereas overly strong noise harms meaningful content information.}

We then perform a grid search over $\{0.1, 0.2, \ldots, 1.0\}$ to tune the weight parameter $\omega_1$, which controls the relative contribution of \textit{interaction-guided denoising} in the overall objective.
Larger values of $\omega_1$ increase the emphasis on noise removal when integrating multimodal embeddings.
Specifically, different domains domains differ in their noise levels and signal structure, they require different degrees of denoising regularization to strike a balance between retaining useful information and suppressing spurious patterns.
For the Cell dataset, the best performance is obtained with a relatively small $\omega_1$, while Sports and Baby perform better with stronger denoising.
Specifically, Fig.~\ref{fig: parameters} (line 2) shows that GTC achieves its peak performance at $\omega_1=0.6$ on Sports, $\omega_1=0.7$ on Baby, and $\omega_1=0.3$ on Cell.
It indicates that \textit{different domains exhibit varying degrees of cross-modal alignment}, and thus requiring different levels of emphasis on denoising process.

We further tune the weight $\omega_2$, which controls the strength of the total correlation objective.
Line 3 of Fig.~\ref{fig: parameters} line 3shows that changing $\omega_2$ noticeably affects the evaluation metrics, indicating the presence of higher-order inter-modal correlations in all three datasets.
Although the response curves differ, each dataset benefits from the total correlation term when $\omega_2$ reaches an appropriate range.
On Sports, performance exhibits volatile but generally improving trends as increases from 0.1 to 0.6, with the highest Recall@20 achieved at $\omega_2=0.6$, after which performance declines as $\omega_2$ grows further.
A similar pattern is observed on Cell, where both Recall@20 and NDCG@20 peak at $\omega_2=0.5$ and then decreases.
Differently, on Baby, the evaluation metrics are maximized at $\omega_2=0.1$ and then decrease with fluctuations as $\omega_2$ increases.
These results suggest that \textit{a moderate emphasis on total correlation helps extract high-order cross-modal signals, whereas an overlarge weight may interfere with other learning objectives}, leading to suboptimal performance.





\vspace{-1mm}
\section{Conclusion}
We introduced GTC to overcome two critical limitations in MMR.
To address the universal feature relevance assumption, GTC employs user-conditional filtering via an interaction-guided diffusion model, which denoises content representations to align with individual user preferences.
To achieve a more complete alignment, GTC maximizes a contrastive lower bound of total correlation across all modalities, capturing the higher-order dependencies that previous studies ignore.
This way, GTC produces item representations that are both personalized and coherently aligned.
We hope the principles of user-conditional generation and holistic alignment will open avenues for future MMR research.

\clearpage

\bibliographystyle{ACM-Reference-Format}
\bibliography{sigir2026}

@article{liu2024multimodal,
  title     = {Multimodal recommender systems: A survey},
  author    = {Liu, Qidong and Hu, Jiaxi and Xiao, Yutian and Zhao, Xiangyu and Gao, Jingtong and Wang, Wanyu and Li, Qing and Tang, Jiliang},
  journal   = {ACM Computing Surveys},
  volume    = {57},
  number    = {2},
  pages     = {1--17},
  year      = {2024},
  publisher = {ACM New York, NY}
}

@article{SMORE,
  title   = {Spectrum-based Modality Representation Fusion Graph Convolutional Network for Multimodal Recommendation},
  author  = {Ong, Rongqing Kenneth and Khong, Andy WH},
  journal = {arXiv preprint arXiv:2412.14978},
  year    = {2024}
}

@inproceedings{liu2022multi,
  title     = {Multi-modal contrastive pre-training for recommendation},
  author    = {Liu, Zhuang and Ma, Yunpu and Schubert, Matthias and Ouyang, Yuanxin and Xiong, Zhang},
  booktitle = {Proceedings of the 2022 International Conference on Multimedia Retrieval},
  pages     = {99--108},
  year      = {2022}
}

@inproceedings{xu2024cmclrec,
  title     = {Cmclrec: Cross-modal contrastive learning for user cold-start sequential recommendation},
  author    = {Xu, Xiaolong and Dong, Hongsheng and Qi, Lianyong and Zhang, Xuyun and Xiang, Haolong and Xia, Xiaoyu and Xu, Yanwei and Dou, Wanchun},
  booktitle = {Proceedings of the 47th International ACM SIGIR Conference on Research and Development in Information Retrieval},
  pages     = {1589--1598},
  year      = {2024}
}

@inproceedings{zhang2020content,
  title     = {Content-collaborative disentanglement representation learning for enhanced recommendation},
  author    = {Zhang, Yin and Zhu, Ziwei and He, Yun and Caverlee, James},
  booktitle = {Proceedings of the 14th ACM Conference on Recommender Systems},
  pages     = {43--52},
  year      = {2020}
}

@article{yang2023diffusion,
  title     = {Diffusion models: A comprehensive survey of methods and applications},
  author    = {Yang, Ling and Zhang, Zhilong and Song, Yang and Hong, Shenda and Xu, Runsheng and Zhao, Yue and Zhang, Wentao and Cui, Bin and Yang, Ming-Hsuan},
  journal   = {ACM Computing Surveys},
  volume    = {56},
  number    = {4},
  pages     = {1--39},
  year      = {2023},
  publisher = {ACM New York, NY, USA}
}

@inproceedings{ronneberger2015u,
  title        = {U-net: Convolutional networks for biomedical image segmentation},
  author       = {Ronneberger, Olaf and Fischer, Philipp and Brox, Thomas},
  booktitle    = {Medical image computing and computer-assisted intervention--MICCAI 2015: 18th international conference, Munich, Germany, October 5-9, 2015, proceedings, part III 18},
  pages        = {234--241},
  year         = {2015},
  organization = {Springer}
}

@article{watanabe1960information,
  title     = {Information theoretical analysis of multivariate correlation},
  author    = {Watanabe, Satosi},
  journal   = {IBM Journal of research and development},
  volume    = {4},
  number    = {1},
  pages     = {66--82},
  year      = {1960},
  publisher = {IBM}
}

@article{saporta2025contrasting,
  title   = {Contrasting with Symile: Simple Model-Agnostic Representation Learning for Unlimited Modalities},
  author  = {Saporta, Adriel and Puli, Aahlad Manas and Goldstein, Mark and Ranganath, Rajesh},
  journal = {Advances in Neural Information Processing Systems},
  volume  = {37},
  pages   = {56919--56957},
  year    = {2025}
}

@article{rendle2012bpr,
  title   = {BPR: Bayesian personalized ranking from implicit feedback},
  author  = {Rendle, Steffen and Freudenthaler, Christoph and Gantner, Zeno and Schmidt-Thieme, Lars},
  journal = {arXiv preprint arXiv:1205.2618},
  year    = {2012}
}

@inproceedings{zhou2023bootstrap,
  author    = {Zhou, Xin and Zhou, Hongyu and Liu, Yong and Zeng, Zhiwei and Miao, Chunyan and Wang, Pengwei and You, Yuan and Jiang, Feijun},
  title     = {Bootstrap Latent Representations for Multi-Modal Recommendation},
  booktitle = {Proceedings of the ACM Web Conference 2023},
  pages     = {845–854},
  year      = {2023}
}

@incollection{zhou2023enhancing,
  title     = {Enhancing dyadic relations with homogeneous graphs for multimodal recommendation},
  author    = {Zhou, Hongyu and Zhou, Xin and Zhang, Lingzi and Shen, Zhiqi},
  booktitle = {ECAI 2023},
  pages     = {3123--3130},
  year      = {2023},
  publisher = {IOS Press}
}

@inproceedings{zhou2023tale,
  title     = {A tale of two graphs: Freezing and denoising graph structures for multimodal recommendation},
  author    = {Zhou, Xin and Shen, Zhiqi},
  booktitle = {Proceedings of the 31st ACM International Conference on Multimedia},
  pages     = {935--943},
  year      = {2023}
}

@inproceedings{wei2020graph,
  title     = {Graph-refined convolutional network for multimedia recommendation with implicit feedback},
  author    = {Wei, Yinwei and Wang, Xiang and Nie, Liqiang and He, Xiangnan and Chua, Tat-Seng},
  booktitle = {Proceedings of the 28th ACM international conference on multimedia},
  pages     = {3541--3549},
  year      = {2020}
}

@inproceedings{zhang2021mining,
  title     = {Mining latent structures for multimedia recommendation},
  author    = {Zhang, Jinghao and Zhu, Yanqiao and Liu, Qiang and Wu, Shu and Wang, Shuhui and Wang, Liang},
  booktitle = {Proceedings of the 29th ACM international conference on multimedia},
  pages     = {3872--3880},
  year      = {2021}
}

@inproceedings{yu2023multi,
  title     = {Multi-view graph convolutional network for multimedia recommendation},
  author    = {Yu, Penghang and Tan, Zhiyi and Lu, Guanming and Bao, Bing-Kun},
  booktitle = {Proceedings of the 31st ACM international conference on multimedia},
  pages     = {6576--6585},
  year      = {2023}
}

@inproceedings{guo2024lgmrec,
  title     = {LGMRec: local and global graph learning for multimodal recommendation},
  author    = {Guo, Zhiqiang and Li, Jianjun and Li, Guohui and Wang, Chaoyang and Shi, Si and Ruan, Bin},
  booktitle = {Proceedings of the AAAI Conference on Artificial Intelligence},
  pages     = {8454--8462},
  year      = {2024}
}

@article{tao2022self,
  title     = {Self-supervised learning for multimedia recommendation},
  author    = {Tao, Zhulin and Liu, Xiaohao and Xia, Yewei and Wang, Xiang and Yang, Lifang and Huang, Xianglin and Chua, Tat-Seng},
  journal   = {IEEE Transactions on Multimedia},
  volume    = {25},
  pages     = {5107--5116},
  year      = {2022},
  publisher = {IEEE}
}

@article{reimers2019sentence,
  title   = {Sentence-bert: Sentence embeddings using siamese bert-networks},
  author  = {Reimers, Nils and Gurevych, Iryna},
  journal = {arXiv preprint arXiv:1908.10084},
  year    = {2019}
}

@inproceedings{wei2019mmgcn,
  title     = {MMGCN: Multi-modal graph convolution network for personalized recommendation of micro-video},
  author    = {Wei, Yinwei and Wang, Xiang and Nie, Liqiang and He, Xiangnan and Hong, Richang and Chua, Tat-Seng},
  booktitle = {Proceedings of the 27th ACM international conference on multimedia},
  pages     = {1437--1445},
  year      = {2019}
}

@article{wang2021dualgnn,
  title     = {Dualgnn: Dual graph neural network for multimedia recommendation},
  author    = {Wang, Qifan and Wei, Yinwei and Yin, Jianhua and Wu, Jianlong and Song, Xuemeng and Nie, Liqiang},
  journal   = {IEEE Transactions on Multimedia},
  volume    = {25},
  pages     = {1074--1084},
  year      = {2021},
  publisher = {IEEE}
}

@inproceedings{kim2022mario,
  title     = {MARIO: modality-aware attention and modality-preserving decoders for multimedia recommendation},
  author    = {Kim, Taeri and Lee, Yeon-Chang and Shin, Kijung and Kim, Sang-Wook},
  booktitle = {Proceedings of the 31st ACM international conference on information \& knowledge management},
  pages     = {993--1002},
  year      = {2022}
}

@inproceedings{cao2022cross,
  title     = {Cross-modal knowledge graph contrastive learning for machine learning method recommendation},
  author    = {Cao, Xianshuai and Shi, Yuliang and Wang, Jihu and Yu, Han and Wang, Xinjun and Yan, Zhongmin},
  booktitle = {Proceedings of the 30th ACM international conference on multimedia},
  pages     = {3694--3702},
  year      = {2022}
}

@article{lin2025contrastive,
  title     = {Contrastive Modality-Disentangled Learning for Multimodal Recommendation},
  author    = {Lin, Xixun and Liu, Rui and Cao, Yanan and Zou, Lixin and Li, Qian and Wu, Yongxuan and Liu, Yang and Yin, Dawei and Xu, Guandong},
  journal   = {ACM Transactions on Information Systems},
  year      = {2025},
  publisher = {ACM New York, NY}
}

@inproceedings{he2016deep,
  title     = {Deep residual learning for image recognition},
  author    = {He, Kaiming and Zhang, Xiangyu and Ren, Shaoqing and Sun, Jian},
  booktitle = {Proceedings of the IEEE conference on computer vision and pattern recognition},
  pages     = {770--778},
  year      = {2016}
}

@inproceedings{cai2025learning,
  title     = {Learning disentangled representation for multi-modal time-series sensing signals},
  author    = {Cai, Ruichu and Jiang, Zhifan and Zheng, Kaitao and Li, Zijian and Chen, Weilin and Chen, Xuexin and Shen, Yifan and Chen, Guangyi and Hao, Zhifeng and Zhang, Kun},
  booktitle = {Proceedings of the ACM on Web Conference 2025},
  pages     = {3247--3266},
  year      = {2025}
}

@article{vaswani2017attention,
  title   = {Attention is all you need},
  author  = {Vaswani, Ashish and Shazeer, Noam and Parmar, Niki and Uszkoreit, Jakob and Jones, Llion and Gomez, Aidan N and Kaiser, {\L}ukasz and Polosukhin, Illia},
  journal = {Advances in neural information processing systems},
  volume  = {30},
  year    = {2017}
}

@inproceedings{du2023distributional,
  title        = {Distributional domain-invariant preference matching for cross-domain recommendation},
  author       = {Du, Jing and Ye, Zesheng and Guo, Bin and Yu, Zhiwen and Yao, Lina},
  booktitle    = {2023 IEEE International Conference on Data Mining (ICDM)},
  pages        = {81--90},
  year         = {2023},
  organization = {IEEE}
}

@inproceedings{he2020lightgcn,
  title     = {Lightgcn: Simplifying and powering graph convolution network for recommendation},
  author    = {He, Xiangnan and Deng, Kuan and Wang, Xiang and Li, Yan and Zhang, Yongdong and Wang, Meng},
  booktitle = {Proceedings of the 43rd International ACM SIGIR conference on research and development in Information Retrieval},
  pages     = {639--648},
  year      = {2020}
}

@article{oord2018representation,
  title   = {Representation learning with contrastive predictive coding},
  author  = {Oord, Aaron van den and Li, Yazhe and Vinyals, Oriol},
  journal = {arXiv preprint arXiv:1807.03748},
  year    = {2018}
}

@inproceedings{yu2025mind,
  title={Mind individual information! principal graph learning for multimedia recommendation},
  author={Yu, Penghang and Tan, Zhiyi and Lu, Guanming and Bao, Bing-Kun},
  booktitle={Proceedings of the AAAI Conference on Artificial Intelligence},
  volume={39},
  number={12},
  pages={13096--13105},
  year={2025}
}

@article{ho2020denoising,
  title={Denoising diffusion probabilistic models},
  author={Ho, Jonathan and Jain, Ajay and Abbeel, Pieter},
  journal={Advances in neural information processing systems},
  volume={33},
  pages={6840--6851},
  year={2020}
}

@article{ma2019learning,
  title={Learning disentangled representations for recommendation},
  author={Ma, Jianxin and Zhou, Chang and Cui, Peng and Yang, Hongxia and Zhu, Wenwu},
  journal={Advances in neural information processing systems},
  volume={32},
  year={2019}
}

@inproceedings{cao2022disencdr,
  title={Disencdr: Learning disentangled representations for cross-domain recommendation},
  author={Cao, Jiangxia and Lin, Xixun and Cong, Xin and Ya, Jing and Liu, Tingwen and Wang, Bin},
  booktitle={Proceedings of the 45th International ACM SIGIR conference on research and development in information retrieval},
  pages={267--277},
  year={2022}
}

@article{wang2022disentangled,
  title={Disentangled representation learning for recommendation},
  author={Wang, Xin and Chen, Hong and Zhou, Yuwei and Ma, Jianxin and Zhu, Wenwu},
  journal={IEEE Transactions on Pattern Analysis and Machine Intelligence},
  volume={45},
  number={1},
  pages={408--424},
  year={2022},
  publisher={IEEE}
}

@inproceedings{an2025beyond,
  title={Beyond whole dialogue modeling: Contextual disentanglement for conversational recommendation},
  author={An, Guojia and Zou, Jie and Wei, Jiwei and Zhang, Chaoning and Sun, Fuming and Yang, Yang},
  booktitle={Proceedings of the 48th International ACM SIGIR Conference on Research and Development in Information Retrieval},
  pages={31--41},
  year={2025}
}

\appendix

\end{document}